# Rastreo muscular móvil usando magnetomicrometría

[†]Cameron R. Taylor[1], [†]Seong Ho Yeon[1],
William H. Clark[2], Ellen G. Clarrissimeaux[1],
Mary Kate O'Donnell[2,3],
[‡*]Thomas J. Roberts[2], [‡*]Hugh M. Herr[1]

[1]K. Lisa Yang Center for Bionics, Massachusetts Institute of Technology, Cambridge, MA, USA. [2]Department of Ecology, Evolution, and Organismal Biology, Brown University, Providence, RI, USA. [3]Department of Biology, Lycoming College, Williamsport, PA, USA. [†]Estos autores comparten la primera autoría. [‡]Estos autores comparten la autoría senior. *Estos autores son los correspondientes.

# Resumen

El tejido muscular es el motor de casi todos los movimientos del reino animal, ya que proporciona fuerza, movilidad y destreza. Las tecnologías para medir el movimiento del tejido muscular, como la sonomicrometría, la fluoromicrometría y el ultrasonido, han avanzado considerablemente la comprensión de la biomecánica. Sin embargo, este campo carece de la capacidad de rastrear el movimiento del tejido muscular en el comportamiento animal fuera del laboratorio. Para abordar este problema, presentamos previamente la magnetomicrometría, un método que utiliza pequeños imanes para rastrear de forma inalámbrica los cambios de longitud del tejido muscular, y validamos la magnetomicrometría mediante pruebas estrechamente controladas *in situ*. En este estudio validamos la precisión de la magnetomicrometría en comparación con la fluoromicrometría usando un modelo de pavo *in vivo* mientras corre libremente. Demostramos el rastreo en tiempo real de la longitud del tejido muscular de los pavos que se mueven libremente ejecutando varias actividades motoras, incluyendo el ascenso y el descenso en rampa, el ascenso y el descenso vertical, y el movimiento libre. Dada la capacidad demostrada de la magnetomicrometría para rastrear el movimiento muscular en animales en un contexto móvil, creemos que esta técnica permitirá nuevas exploraciones científicas y una mejor comprensión de la función muscular.

PALABRAS CLAVE

biomecánica, rastreo muscular, tecnología implantable,
tecnología vestible, rastreo del movimiento, magnetomicrometría, imanes, rastreo magnético

# Introducción

Las mediciones de la longitud muscular han impulsado descubrimientos importantes en la biomecánica del movimiento (Fowler et al. 1993), informado modelos de control motor (Roberts et al. 1997; Prilutsky et al. 1996) y proporcionado estrategias para el diseño de prótesis y robots (Eilenberg et al. 2010). Durante décadas, la sonomicrometría (SM) ha informado sobre cómo se mueven los músculos, proporcionando una alta precisión (70 µm de resolución) y un gran ancho de banda (>250 Hz) (Griffiths 1987). La fluoromicrometría (FM) amplió el conjunto de herramientas de rastreo muscular, permitiendo una alta precisión (90 µm de precisión) y un gran ancho de banda (>250 Hz) para el rastreo de un gran número de marcadores (Brainerd et al. 2010; Camp et al. 2016). Además, el ultrasonido basado en imágenes (U/S) añadió la capacidad de rastrear de forma no invasiva las geometrías musculares (Fukunaga et al. 2001; Sikdar et al. 2014; Clark y Franz 2021).

Sin embargo, la recolección de mediciones directas de la longitud muscular en entornos naturales sigue siendo inviable, por lo que la estimación indirecta de la longitud muscular se sigue utilizando para observar los movimientos naturales. Por ejemplo, las longitudes musculares se estiman utilizando los ángulos de las articulaciones mediante modelos biofísicos (Delp et al. 2007). Estas aproximaciones se utilizan debido a limitaciones de las técnicas actuales de detección del movimiento muscular, las cuales necesitan conexiones alámbricas o equipos voluminosos. Tanto el SM como el U/S requieren conexiones alámbricas con equipo voluminoso para la detección (Biewener et al. 1998; Clark y Franz 2021). El SM requiere cirugía avanzada y cables percutáneos. Y, aunque la FM no requiere una conexión alámbrica, está limitada a un volumen del tamaño de un balón de fútbol aproximadamente, requiere un equipo del tamaño de una habitación pequeña, y está limitada en el tiempo debido a las restricciones térmicas y a la exposición del sujeto a la radiación. (Brainerd et al. 2010).

Las tecnologías actuales de rastreo de la longitud del músculo también requieren un tiempo considerable de posprocesamiento, lo que dificulta su uso en estudios longitudinales. El SM requiere tener en cuenta y filtrar artefactos como los errores de disparo (Marsh 2016), la FM requiere el etiquetado de puntos en imágenes estereoscópicas (Brainerd et al. 2010), y el U/S requiere el etiquetado de fascículos (Van Hooren et al. 2020), los cuales requieren al menos algún tipo de procesamiento manual. Aunque las técnicas de aprendizaje automático han demostrado su potencial para el rastreo automático de la longitud de los fascículos a partir de imágenes de ultrasonido, la falta actual de fiabilidad en el rastreo de las mediciones de actividad cruzada ($R^2$ =0.05 para el entrenamiento de actividad cruzada de un solo sujeto de una máquina de vector soporte) impide que dicha estrategia sea aplicable para detectar la longitud de los fascículos durante el movimiento natural (Rosa et al. 2021).

Los investigadores necesitan una plataforma de detección que pueda funcionar libremente en entornos naturales sin conexiones alámbricas, detectando toda la gama dinámica del movimiento muscular en su contexto. Para satisfacer esta necesidad, hemos desarrollado la magnetomicrometría (MM), una estrategia mínimamente invasiva para el rastreo muscular móvil en tiempo real. La MM utiliza un conjunto de sensores de campo magnético para localizar y calcular la distancia entre dos imanes implantados con un retraso de menos de un milisegundo. Esta distancia proporciona una medida de longitud del tejido muscular entre los imanes implantados. MM permite el registro continuo durante un intervalo de recolección indefinido que se extiende a lo largo de horas, con potencial para ser usado continuamente a lo largo de días, semanas o años.

En trabajos anteriores, validamos el concepto de la MM en comparación con la FM mediante pruebas *in situ* muy controladas (Taylor et al. 2021). Sin embargo, no se ha demostrado empíricamente que la técnica de MM sea robusta para el registro durante la locomoción libre. En el presente estudio abordamos esta pregunta. Investigamos la solidez de la MM durante la actividad libre que presenta artefactos en los tejidos blandos (es decir, el movimiento de los sensores del campo magnético en relación con el músculo) y cambios en la orientación relativa del campo magnético ambiental.

Aquí presentamos la MM como una estrategia robusta, práctica y eficaz para medir la longitud del tejido muscular en un modelo animal de movimiento libre. En primer lugar, aplicamos esta técnica a pavos que corren en una cinta de correr, y comparamos la MM con la FM para determinar la precisión del método. A continuación, investigamos el uso de la MM para rastrear la longitud del tejido muscular en animales que se mueven libremente durante el ascenso y el descenso en rampa, el ascenso y el descenso vertical, y el movimiento libre. Nuestra hipótesis es que la longitud del tejido muscular puede ser rastreada durante el movimiento libre a través de la MM con una precisión submilimétrica y una fuerte correlación ($R^2 > 0.5$) con la MM. La validación de esta herramienta en un contexto móvil permite el rastreo y la investigación de la fisiología muscular en entornos anteriormente inaccesibles para los investigadores de biomecánica.

# Resultados

*Validación de la exactitud de la magnetomicrometría frente a la fluoromicrometría*

Para verificar la precisión del rastreo de la MM durante la actividad libre, se realizó un rastreo de los pares de pequeños imanes implantados en los músculos gastrocnemios de los pavos (pata derecha, tres pavos) utilizando tanto la MM como la FM mientras los pavos caminaban y corrían a múltiples velocidades en una cinta de correr (véase la Figura 1 para la configuración y los resultados del rastreo, y la Figura Suplementaria 1 para un escaneo en 3D de la serie de sensores usado para la MM).

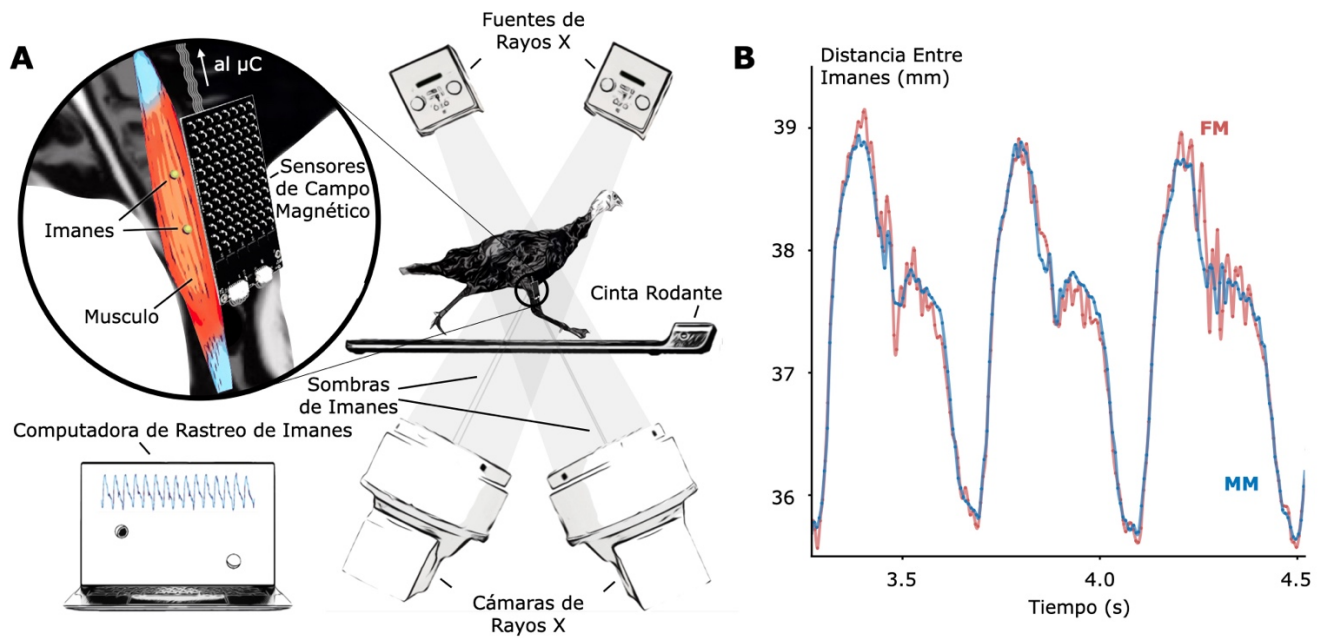

***Figura 1: Validación del rastreo muscular móvil mediante magnetomicrometría.*** *(A) Una serie de sensores de campo magnético en la superficie de la pierna rastrea las posiciones de dos imanes implantados en el músculo. Un microcontrolador (μC) en las plumas del pavo transmite de forma inalámbrica los datos del campo magnético a una computadora de rastreo de imanes que calcula y muestra la señal de la magnetomicrometría (MM) en tiempo real. Los pavos caminaron y corrieron en una cinta de correr mientras las cámaras de vídeo de rayos X grababan los datos sincronizados con la fluoromicrometría (FM) para su procesamiento posterior. **(B)** Comparación de la MM (azul) con la FM (rojo) para validar la precisión de la MM. Estos resultados representativos durante la marcha en carrera muestran la precisión submilimétrica de la MM durante el rastreo de la longitud muscular móvil.*

Comparamos las distancias medidas por MM entre las posiciones de los pequeños imanes con sus distancias medidas por FM para evaluar la precisión durante la actividad en la cinta rodante (véase la Figura 2). Los coeficientes de determinación (valores de $R^2$) entre la MM y la FM fueron 0.952, 0.860 y 0.967 para las aves A, B y C, respectivamente (véase también la Figura Suplementaria 2). Las diferencias entre la MM y la FM fueron de -0.099 ± 0.186 mm, -0.526 ± 0.298 mm y -0.546 ± 0.184 mm para las aves A, B y C, respectivamente (véase la Figura Suplementaria 3).

Para determinar la fiabilidad específica del estudio del procesamiento manual de la FM (etiquetado de la posición del marcador en los datos de vídeo de rayos X), se volvieron a etiquetar manualmente diez ciclos de marcha de los datos brutos de la FM de forma independiente tres veces para un ave a una velocidad. En estos tres etiquetados para estos diez ciclos de marcha, el procesamiento manual de la FM fue consistente con una desviación estándar de 0.098 mm (véase la Figura Suplementaria 4 para más detalles).

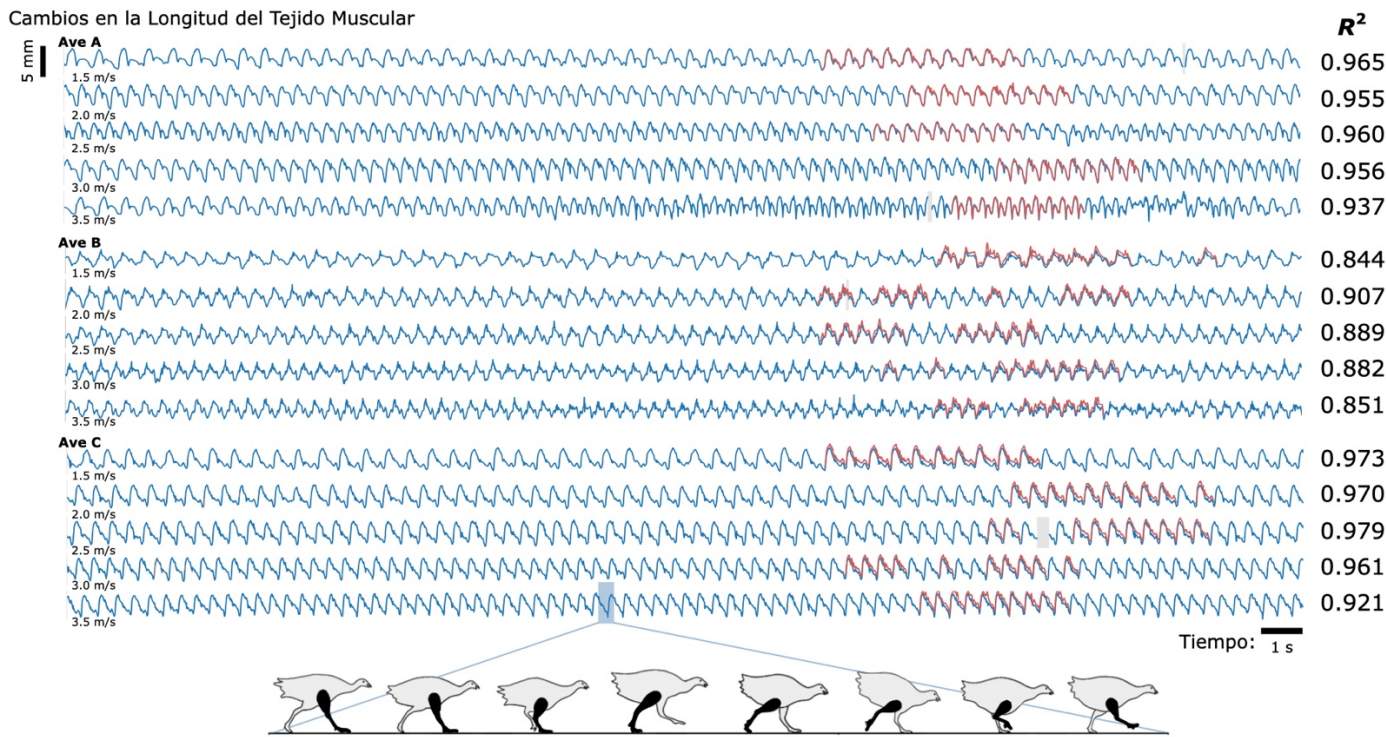

*Figura 2: Rastreo muscular móvil durante la carrera en cinta de correr: La magnetomicrometría en comparación con la fluoromicrometría. Cambios en la longitud del tejido muscular medidos por la MM (azul) y la FM (rojo) para tres pavos a cinco velocidades (se muestran 30 s para cada velocidad). La columna de la derecha de los gráficos da los coeficientes de determinación ($R^2$) entre la magnetomicrometría y la fluoromicrometría correspondientes a cada pavo y velocidad. Las brechas en los datos de la fluoromicrometría se deben a la selección por parte del investigador de ciclos de marcha completos durante los cuales los dos imanes eran visibles en ambas imágenes de rayos X. Las brechas en los datos de magnetomicrometría (en gris) se deben a las caídas de paquetes durante la transmisión inalámbrica de las señales del campo magnético a la computadora de rastreo (brechas por debajo de 50 ms interpoladas en gris, brechas por encima de 50 ms resaltadas en gris). El diagrama de la marcha del pavo debajo de los gráficos muestra las correspondientes fases de la marcha durante un ciclo de la misma.*

Los retrasos de rastreo por la MM en el percentil 99 fueron de 0.698 ms, 0.690 ms y 0.664 ms para las aves A, B y C, respectivamente (véase también la Figura Suplementaria 5), y los datos de MM no requirieron ningún posprocesamiento. Por el contrario, el posprocesamiento de los datos de la FM en distancias de marcador-a-marcador requirió aproximadamente 84 horas de procesamiento manual repartidas en varios meses.

*Rastreo muscular móvil en diversas actividades*

Para investigar la viabilidad del uso de la MM durante un movimiento dinámico y natural, construimos una serie de obstáculos para que los pavos los recorrieran. En concreto, proporcionamos a los pavos dos rampas inclinadas (10° y 18°, véase la Figura 3) y tres cambios de elevación vertical (20 cm, 41 cm y 61 cm, véase la Figura 4). Dado que el propósito de estas actividades era explorar la gama de movimientos dinámicos que se podían capturar, no entrenamos a las aves para que navegaran por las rampas o los cambios de elevación verticales de forma repetitiva, por lo que se espera que haya variabilidad en las tareas repetidas.

Para ver un vídeo de uno de los pavos (Ave A) navegando por todos estos obstáculos con datos de MM en tiempo real, véase la Película Suplementaria 1.

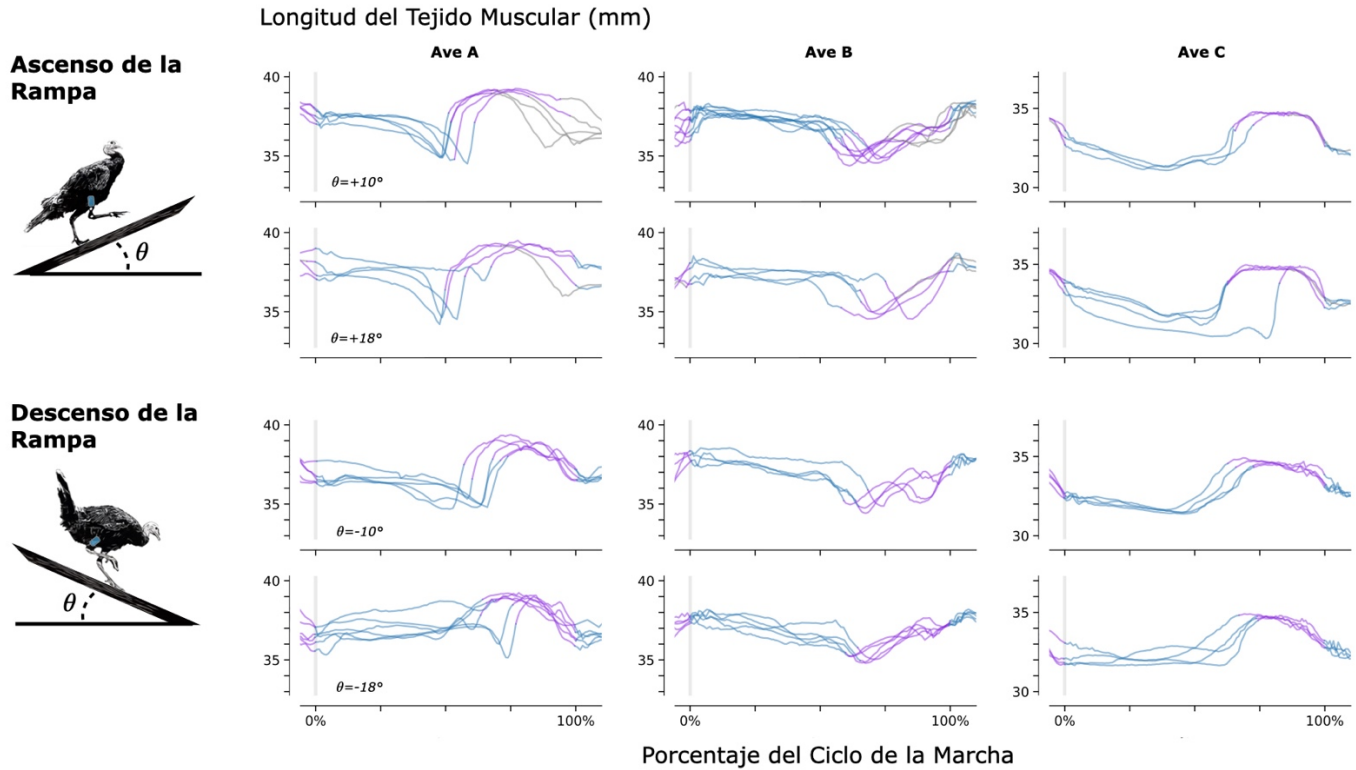

*Figura 3: **Longitud del tejido muscular durante el ascenso y descenso no sincronizado de la rampa.*** *Utilizamos la magnetomicrometría para rastrear la longitud del tejido muscular durante el ascenso y el descenso de la rampa en dos pendientes para las tres aves. Los datos de cada ave y cada pendiente están sincronizados en el golpe de la pata derecha (indicado por la línea gris vertical) y normalizados de golpe de pata a golpe de pata. La variabilidad entre las curvas refleja la variabilidad del ciclo de la marcha durante la navegación en rampa sin entrenamiento. La longitud del tejido muscular se representa en azul para el apoyo de la pierna derecha, en morado para el balanceo de la pierna derecha y en gris cuando el vídeo no permitió etiquetar la fase de la marcha. Registramos al menos tres ciclos de marcha de cada actividad por cada ave.*

Taylor et al. Rastreo Muscular Móvil

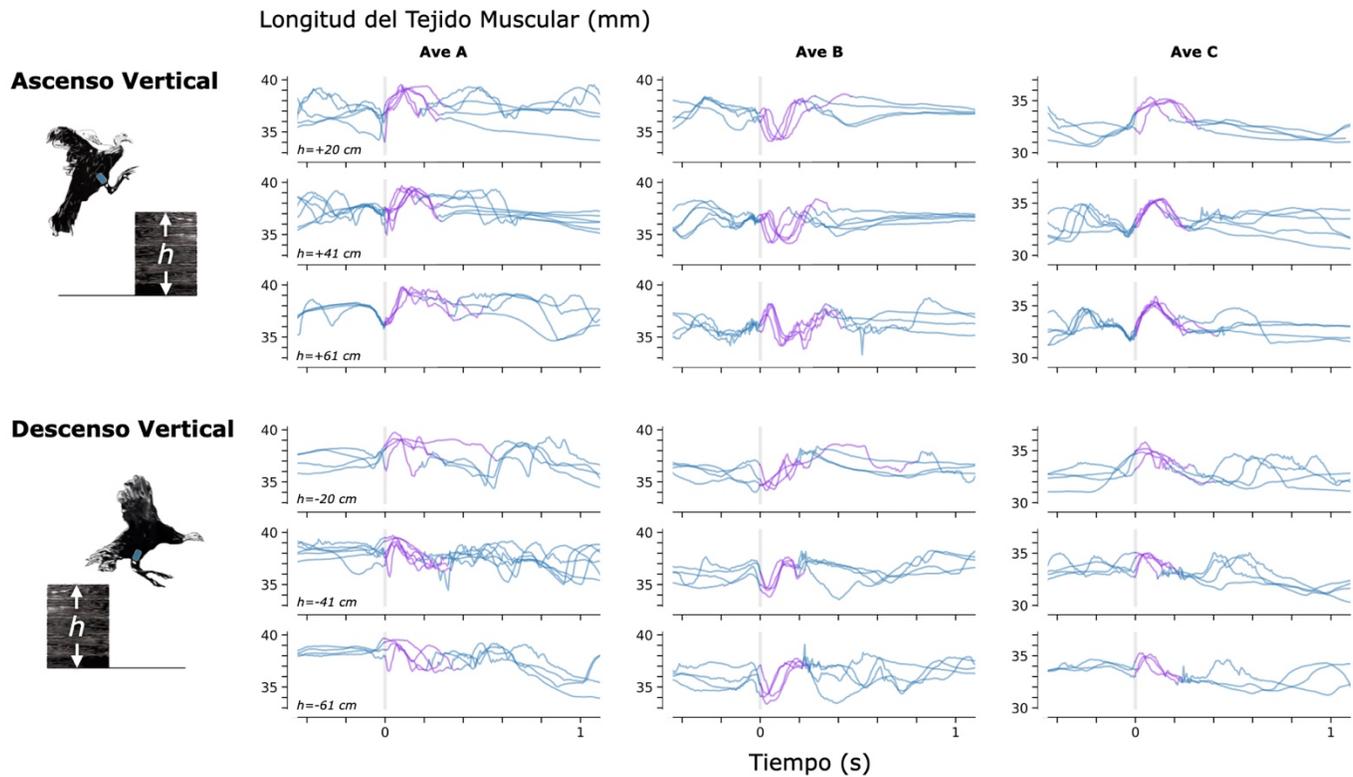

*Figura 4: Longitud del tejido muscular durante el ascenso y descenso vertical no sincronizado.* Utilizamos la magnetomicrometría para seguir la longitud del tejido muscular durante el ascenso y el descenso vertical a tres alturas para las tres aves. Los datos de cada ave y cada altura están sincronizados en el momento de despegue de la pata derecha (inicio de la fase aérea, indicado con la línea gris vertical). La variabilidad entre las curvas refleja la variabilidad del movimiento durante el ascenso y el descenso vertical sin entrenamiento. La longitud del tejido muscular durante el contacto con el suelo se representa en azul, y la longitud del tejido muscular durante la fase aérea se representa en morado. Se muestran todos los datos, incluidos los escenarios en los que se produjo un aleteo significativo durante el salto hacia arriba o hacia abajo. Capturamos al menos tres grabaciones de cada actividad por cada ave.

Para validar aún más la precisión de la MM utilizada durante la navegación por rampas y cambios de elevación verticales, analizamos los datos de rastreo de los imanes de estas actividades para encontrar el rango de las posiciones tridimensionales rastreadas de los imanes (véase la Figura Suplementaria 6). A continuación, fijamos dos imanes a 40 mm de distancia, validamos la distancia entre ellos mediante la FM ($40.000 \pm 0.017$ mm) y barrimos este par validado de imanes por la FM bajo la serie de sensores de la MM a través de un volumen que superaba estos rangos (véase la Figura Suplementaria 7). Controlamos las desviaciones de 40 mm en la señal de MM durante estas pruebas de banco y encontramos un error de percentil 99 ($e_{99\%}$) de 1.000 mm (redondeado al micrómetro más cercano).

Por último, para explorar si el rastreo muscular móvil a través de la MM es viable en un contexto de itinerancia libre, rastreamos la longitud del tejido muscular mientras un pavo (Ave A) vagaba libremente por su recinto. Los resultados de esta recolección de datos se muestran en la Figura 5.

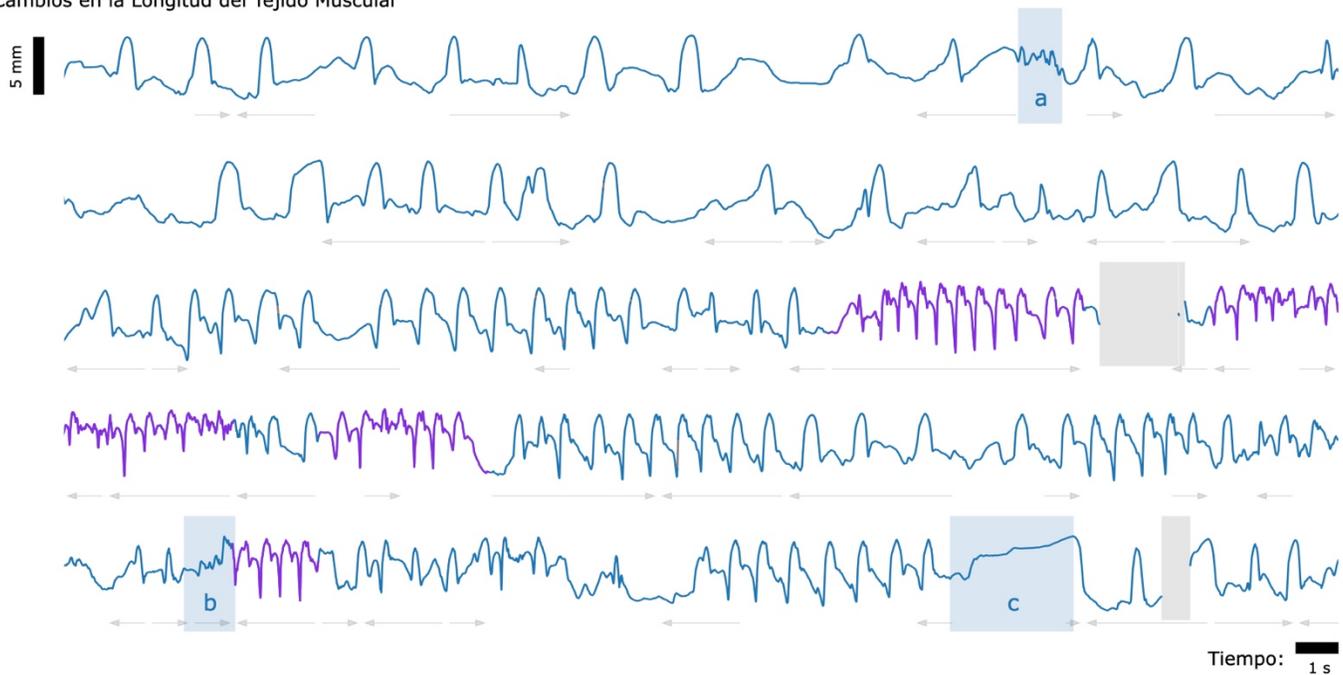

*Figura 5: Longitud del tejido muscular durante el movimiento de movimiento libre. Los datos de la magnetomicrometría se recogieron de forma continua durante 150 segundos durante la actividad de movimiento libre. La longitud del tejido muscular se representa en azul durante la bipedestación y la marcha, y en morado durante la carrera. Las regiones resaltadas en azul indican la longitud del tejido muscular durante (a) el erizado de plumas, (b) el salto y (c) el equilibrio sobre una pata. Las flechas grises indican cuándo el pavo giraba a la izquierda (flechas de la izquierda) o a la derecha (flechas de la derecha). Las brechas debidas a las caídas de paquetes de transmisión inalámbrica se muestran en gris, como se describe en la Figura 2.*

## Discusión

Descubrimos que la MM permite el rastreo de la longitud del tejido muscular móvil con una alta correlación con la MM ($R^2$ de 0.952, 0.860 y 0.967 para las aves A, B y C, respectivamente) y una precisión submilimétrica (media más o menos la desviación típica de -0.099 ± 0.186 mm, -0.526 ± 0.298 mm y -0.546 ± 0.184 mm para las aves A, B y C, respectivamente). Estos hallazgos permiten el rastreo y la investigación del comportamiento contráctil del músculo en entornos antes inaccesibles para los investigadores de biomecánica.

*Validación de la precisión*

El estándar que utilizamos aquí para evaluar la exactitud del rastreo de la longitud del músculo utilizando la MM fue la FM. En el caso de los imanes implantados superficialmente en los músculos (a profundidades inferiores a 2 cm), la MM presenta menos ruido que la MM, pero la MM tiene la ventaja de una mayor precisión, especialmente a mayores profundidades del tejido (las profundidades de rastreo en este estudio oscilaron entre 11.2 mm y 26.6 mm). De hecho, nuestras pruebas mostraron que para los marcadores no ocultos que se mueven a través del volumen de rayos X, la FM fue precisa hasta 0.030 mm. Sin embargo, observamos que el ruido de rastreo de los marcadores fue un reto para la FM en este estudio en particular, debido al uso de un animal grande y a la presencia de hardware (el conjunto de detección de la MM) que regularmente oscurecía los marcadores durante el rastreo. Estos factores dieron lugar a un ruido de etiquetado manual sustancial en la señal de la FM de 0.098 mm, en lugar del ruido de 0.030 mm que encontramos en nuestra prueba de precisión de la FM, lo que afecta a las desviaciones típicas de precisión indicadas anteriormente. Si se tiene en cuenta este ruido de etiquetado manual, se obtienen desviaciones típicas de precisión ajustadas de 0.158, 0.281 y 0.156 mm, para las aves A, B y C, respectivamente (véase la Figura Suplementaria 3).

Las restricciones del volumen de detección hacen que la FM sea poco práctica durante la actividad en terreno variable de los animales grandes, por lo que realizamos pruebas de precisión retrospectivas en el banco para validar aún más los datos de la MM recogidos durante la navegación de las rampas y los cambios de elevación vertical (véase la Figura Suplementaria 7). Los artefactos de los tejidos blandos durante los movimientos dinámicos, como la deformación o el movimiento de los tejidos, dan lugar a cambios de profundidad y posición de los imanes con relación a la serie de sensores de la MM. En las pruebas de banco se investigó la precisión de las mediciones de la MM en toda la gama de profundidades y posiciones de los imanes que observamos durante esas actividades (véase la Figura Suplementaria 6). El error que observamos en las pruebas de banco ($e_{99\%} < 1$ mm) fue aceptable en comparación con la magnitud de las contracciones musculares que observamos durante la actividad en terreno variable (la magnitud media de la señal de la MM fue de 4.5 mm de pico a pico). Esto sugiere que la MM rastreó con solidez las longitudes del tejido muscular durante las actividades en terreno variable, a pesar de los artefactos de los tejidos blandos que pudieron producirse durante los movimientos dinámicos que exigían dichas actividades. Sin embargo, estas pruebas ponen de manifiesto la importancia de la colocación del sensor. Se consigue una mayor precisión cuando el conjunto de sensores de la MM está correctamente colocado, centrado sobre los imanes. En teoría, la MM con una detección perfecta del campo magnético no se vería afectada por el movimiento de los sensores en relación con los imanes implantados, pero los errores observados sugieren que los sensores no son lineales. La compensación de la no linealidad del rastreo del imán (por ejemplo, mediante la calibración del sensor o las geometrías tridimensionales del mismo) es, por tanto, un área importante para investigación futura. Mientras tanto, en trabajo futuro, una ventaja sería disponer de una serie de sensores más grande con una cobertura más amplia, para mitigar la necesidad de una colocación cuidadosa de la serie de sensores.

*Campos magnéticos ambientales*

La compensación de perturbaciones magnéticas basada en software que empleamos aquí (Taylor et al. 2019) fue suficiente para compensar los campos magnéticos ambientales durante el rastreo muscular móvil en presencia de grandes mesas de elevación hidráulicas ferromagnéticas, un motor de cinta rodante grande y activo, y una sala llena de equipos de rayos X activos. Sin embargo, nuestra estrategia de compensación de perturbaciones uniformes puede ser insuficiente para la situación excepcional en la que un gran objeto ferromagnético está inmediatamente adyacente (a unos pocos centímetros) al músculo rastreado. Por lo tanto, la compensación basada en software para campos magnéticos ambientales espacialmente no uniformes puede ser una dirección valiosa para el trabajo futuro para ampliar la robustez de la MM a ese escenario potencial. Como alternativa, se podría utilizar un blindaje ferromagnético para realizar físicamente la compensación de las perturbaciones (Tarantino et al. 2017), pero el blindaje tendría que estar lo suficientemente lejos para evitar que actuara como un espejo magnético, creando imanes "de imágenes" que también tendrían que ser rastreados (Hammond 1960). Además, un blindaje eficaz tendría que ser lo suficientemente grueso como para redirigir la mayoría o todas las perturbaciones del campo magnético, lo que supone un compromiso entre el peso y la eficacia del blindaje.

*Gama de comportamientos*

Las Figuras 3, 4 y 5 ofrecen una muestra de la gama de comportamientos que pueden seguirse mediante la magnetomicrometría. No se buscó, ni se esperó, ni se deseó la consistencia de las curvas. Más bien, preservamos intencionalmente los eventos anómalos en esos datos, como los aleteos únicos o múltiples durante el ascenso y el descenso vertical y la velocidad variable durante la navegación en rampa, para explorar la gama de actividades motoras durante las cuales podíamos rastrear la actividad muscular.

*Aplicaciones*

La MM tiene el potencial de funcionar en todas las escalas (véase la Figura 6), desde la capacidad de rastrear el movimiento de todo el cuerpo y de los músculos de organismos pequeños hasta la capacidad de rastrear grandes imanes implantados en profundidad en modelos de animales grandes. Matemáticamente, si el número de sensores es fijo y todas las dimensiones del sistema se escalan, el error como porcentaje de la excursión escalada del imán permanecerá sin cambios (Taylor et al. 2019). Sin embargo, en principio se pueden utilizar series de sensores más grandes cuando se rastrean animales muy pequeños o muy grandes, lo que da lugar a un aumento de la precisión del rastreo en esos extremos. Por ejemplo, cuando se rastrea el tejido muscular de animales pequeños, se pueden incrustar sensores adicionales en el entorno del animal, y cuando se rastrea el tejido muscular de animales grandes, el aumento del tamaño del animal permite montar sensores adicionales en el animal. De este modo, los sistemas de rastreo de imanes, cuando son específicamente diseñados para cada contexto, pueden aprovechar las geometrías únicas que ofrece cada escala.

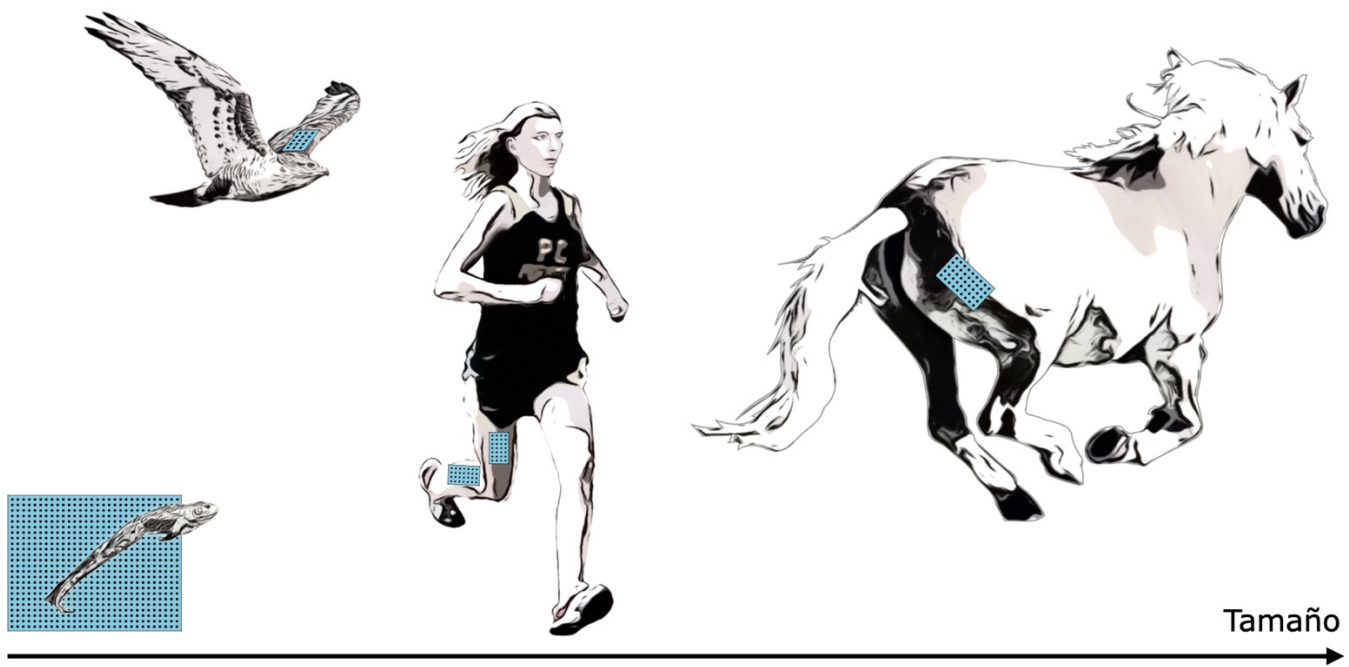

*Figura 6: Rastreo de los músculos a distintas escalas. Cambiando el tamaño de la serie de sensores del campo magnético, podemos rastrear la distancia entre los imanes a distancias más cercanas o más lejanas, lo que nos permite, en principio, rastrear los tejidos musculares a una serie de escalas, incluyendo ranas, halcones, personas, caballos u otros animales. En el caso de los animales pequeños, como la rana que se muestra en la parte inferior izquierda, una serie de sensores fijados debajo o al lado del animal podría rastrear tanto la posición del animal como la longitud del tejido muscular.*

La MM no sólo puede utilizarse a través de escalas de tamaño, sino también a través de escalas de tiempo. Dado que no es necesario el posprocesamiento, los datos de la MM pueden recogerse de forma continua, lo que permite realizar estudios longitudinales, incluyendo investigaciones sobre mecanismos como la degradación neuronal o la plasticidad a lo largo del tiempo.

Las señales de longitud y velocidad muscular de la MM son diferentes y complementarias a las señales de la electromiografía (EMG). Mientras que la EMG proporciona una medida de la activación muscular, que resulta más directamente de las órdenes neuronales, la longitud y la velocidad del músculo dan información sobre la forma del músculo, que a su vez puede perfeccionar nuestra comprensión de la fisiología muscular durante una tarea de movimiento determinada. De hecho, la combinación de la MM y la EMG permitirá aumentar la comprensión fisiológica en nuevos contextos en los que los animales se encuentren en sus entornos naturales.

En un trabajo paralelo, también demostramos la viabilidad de los implantes de imanes para uso humano, verificando la comodidad, la ausencia de migración del implante y la biocompatibilidad (Taylor, Clark, et al. 2022). Debido a la naturaleza móvil de la MM, esta técnica tiene aplicaciones en el control protésico y exoesquelético. En su implementación más sencilla, un controlador de motor podría controlar directamente una articulación robótica utilizando la distancia entre dos imanes de cada músculo de un par flexor-extensor. Sin embargo, la capacidad de la MM de rastrear otros músculos y de trabajar en combinación con la EMG permite una serie de nuevas estrategias para la interfaz hombre-máquina.

*Limitaciones*

En este estudio, implantamos los imanes a una distancia de aproximadamente 3.5 cm entre sí, basándonos en trabajos anteriores (Taylor et al. 2021), para asegurar que los imanes no migraran unos hacia otros. Si se utilizan implantes de imanes más pequeños o más grandes (o con formas diferentes, por ejemplo, en un modelo animal más pequeño o más grande), el efecto de la distancia de separación en la estabilidad contra la migración tendría que volver a investigarse para los diferentes tamaños (y las diferentes fuerzas de magnetización) de los implantes.

Para las pruebas de validación de precisión en el laboratorio, asumimos que las posiciones de los imanes rastreados eran una buena aproximación a las posiciones reales de los imanes. Utilizamos la información de rastreo de la posición del imán a partir de los datos de la MM de terreno variable para determinar los límites del volumen a probar. A continuación, al inicio de las pruebas, utilizamos las posiciones de los imanes rastreados para localizar la posición centrada y de profundidad mínima (la posición más cercana dentro del rango de escala completa de los sensores), y luego utilizamos bloques de dimensiones conocidas para barrer la profundidad del tejido emulado en el banco y hacer cumplir los límites del volumen. Teniendo en cuenta que la MM tenía una precisión submilimétrica en todo el volumen, consideramos que estas suposiciones eran razonables para estas pruebas.

Al igual que para cualquier rastreo del tejido muscular, la ubicación de los dispositivos implantados de rastreo determinará la longitud medida. Para los estudios en los que el objetivo es relacionar los cambios de longitud medidos con las propiedades contráctiles del músculo (por ejemplo, las relaciones longitud-tensión o fuerza-velocidad), es esencial que los marcadores estén alineados a lo largo del eje del fascículo. En el presente estudio, incrustamos los imanes en lugares de los músculos del pavo que garantizaran que los imanes permanecieran en su lugar durante un período de meses, y a profundidades que fueran favorables para la función del sensor. Por lo tanto, los patrones de cambio de longitud no representan directamente los patrones de cambio de longitud de los fascículos musculares y pueden estar influidos significativamente por los cambios dinámicos en la arquitectura muscular durante la contracción. Esto se refleja en los cambios opuestos de la longitud del tejido muscular observados durante la fase de balanceo del ave B en relación con las aves A y C durante la navegación en rampa (véase la Figura 3). La MM, la FM y la SM adolecen de este mismo problema y, por lo tanto, para cualquiera de estas técnicas se justifica una cuidadosa colocación quirúrgica.

*Mejoras en la detección*

El conjunto de componentes electrónicos para la MM es inmediatamente actualizable a medida que se desarrollan nuevos estándares industriales. El sistema de rastreo se beneficia de los avances mundiales en sensores de campo magnético de bajo coste debido a la fabricación generalizada de unidades de medición inercial para dispositivos como teléfonos móviles, controladores de videojuegos y vehículos autónomos. Las mejoras continuas en los sensores de campo magnético, los condensadores y los microcontroladores provocarán mejoras directas en la precisión, la eficacia y la velocidad del sistema de rastreo y permitirán el rastreo de implantes aún más pequeños a mayor profundidad.

*Resumen*

Aquí, demostramos el uso de la MM para el rastreo muscular móvil. Validamos, en comparación con la FM, la precisión submilimétrica de la MM en un modelo de pavo despierto y activo ($R^2 \geq 0.860$, $\mu \leq 0.546$ mm, $\sigma \leq 0.298$ mm) con un retraso de tiempo de cálculo en tiempo real de menos de un milisegundo ($\eta_{0.99} \leq 0.698$ ms). Demostramos además el uso de la MM en el rastreo muscular móvil durante el ascenso y el descenso en rampa, el ascenso y el descenso vertical, y el movimiento libre. Estos resultados alientan el uso de la MM en futuras investigaciones biomecánicas, así como en el control protésico y exoesquelético. Esperamos que la MM permita una variedad de nuevos experimentos y tecnologías, y esperamos con interés el desarrollo y la aplicación de esta tecnología.

# Métodos

Todos los experimentos con animales fueron aprobados por los Comités Institucionales de Cuidado y Uso de Animales de la Universidad de Brown y del Instituto Tecnológico de Massachusetts. Los pavos silvestres (*Meleagris gallopavo*, hembra adulta) se obtuvieron de criadores locales y se mantuvieron en las instalaciones de cuidado de animales de la Universidad de Brown con una dieta de agua y comida para aves ad libitum. En este estudio se utilizaron tres animales.

*Procedimiento quirúrgico*

Se implantaron un par de imanes recubiertos de parileno de 3 mm de diámetro (N48SH) en el músculo gastrocnemio lateral derecho de cada pavo, con el objetivo de una distancia de separación de 3.5 cm entre imanes. Para más detalles sobre el procedimiento quirúrgico y los implantes, véase Taylor et al. 2022. Se dio un periodo de recuperación de un mes antes de comenzar la recolección de datos.

*Magnetomicrometría*

Para este estudio, diseñamos una placa de sensores de campo magnético personalizada (véase la Figura 7). La placa de sensores estaba equipada con 96 sensores de campo magnético (LIS3MDL, STMicroelectronics) separados 5.08 mm en una cuadrícula de 8 por 12. Cada sensor estaba equipado con condensadores no magnéticos (VJ1206Y105KCXAT y VJ0603Y104KCXAT, Vishay). Siete multiplexores digitales en el conjunto de sensores permitían la multiplexación en el dominio del tiempo (un 74HC138BQ,115 multiplexado en seis 74HC154BQ,118, Nexperia) a través de una conexión por cable. La placa de sensores se conectó a través de una placa adaptadora personalizada a un sistema integrado de microcontrolador inalámbrico disponible en el mercado (microcontrolador WiFi Feather M0, Adafruit), que se alimentó con una batería de polímero de iones de litio (3.7 V, 1800 mA·h, 29 g). El microcontrolador muestreaba las señales del campo magnético a 155 Hz y las transmitía de forma inalámbrica a la computadora de rastreo de los imanes a través de un router WiFi (Nighthawk R6900P, Netgear). El algoritmo de rastreo se ejecutó en tiempo real en la computadora de rastreo de imanes, un portátil Dell Precision 5550 (sistema operativo Ubuntu 20.04) con 64 GB de memoria de acceso aleatorio y un procesador Intel i7 de 8 núcleos, que funcionaba a 2.30 GHz. El algoritmo de rastreo utilizado, incluida la estrategia de compensación de perturbaciones, se detalla en el trabajo anterior (Taylor et al. 2019).

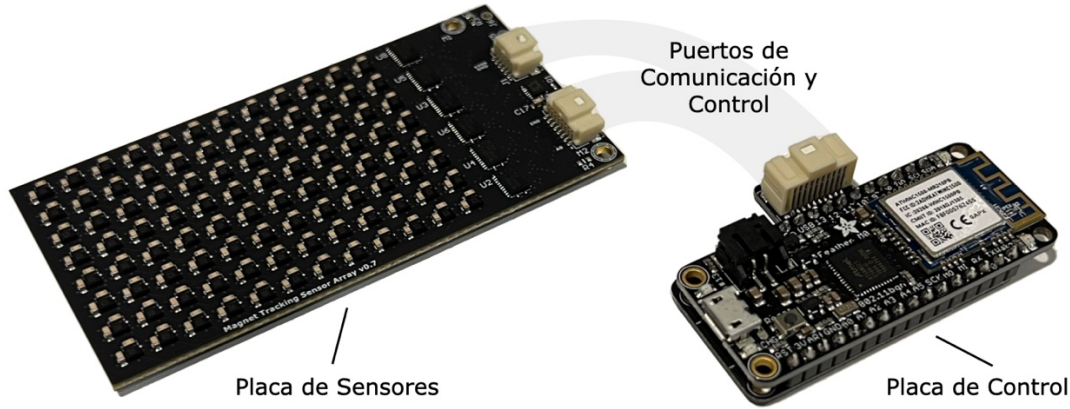

*Figura 7: Sistema integrado de la magnetomicrometría. Fabricamos una placa de sensores personalizada (izquierda) y una placa de control personalizada (derecha) para este estudio. La placa de sensores contiene la serie de sensores de la magnetomicrometría, que consiste en 96 sensores de campo magnético dispuestos con una separación de 5.08 mm. Los multiplexores digitales de la placa de sensores permiten la multiplexación en el dominio del tiempo, lo que permite que un único microcontrolador de la placa de control se comunique con todos los sensores de campo magnético de la placa de sensores y los controle. La placa de control combina los datos de la placa de sensores y los transmite de forma inalámbrica a la computadora de rastreo de imanes. La placa de sensores y la de control pesan 24 g y 12 g, respectivamente.*

Fijamos el conjunto de sensores a la extremidad utilizando la película adhesiva Opsite Flexifix (Smith & Nephew). Para fijar dicho conjunto, primero aplicamos una capa base de la película adhesiva a la piel. A continuación, colocamos la placa de sensores sobre la pierna y envolvimos la película adhesiva alrededor de la placa de sensores y la pierna. Para mantener una distancia mínima suficiente entre los imanes y la placa de sensores, colocamos capas de espuma entre el adhesivo base y la placa de sensores. A continuación, fijamos la placa de control y la batería dentro de las plumas traseras del pavo.

*Validación de la exactitud de la magnetomicrometría en comparación con la fluoromicrometría*

Utilizamos la instalación XROMM de la Fundación W. M. Keck en la Universidad de Brown (Brainerd et al. 2010) para realizar la FM. Recogimos el vídeo de rayos X de dos haces de rayos X que se cruzaban y que estaban orientados a 51 grados el uno del otro. Montamos una cinta de correr con una base de madera (TM145, Horizon Fitness) entre las cámaras de rayos X y las fuentes de rayos X y construimos un recinto sobre la cinta de correr con una pared móvil para colocar las aves dentro de la ventana de captura.

Los pavos caminaron y corrieron a cinco velocidades (1.5, 2.0, 2.5, 3.0 y 3.5 m/s) en un orden aleatorio hasta que fueron visibles al menos diez ciclos de marcha dentro del volumen de captura de la FM para cada velocidad. Se recogieron datos de la FM a 155 Hz.

La sincronización del tiempo se realizó a través de una conexión por cable coaxial desde la FM a una placa de desarrollo de microcontroladores de venta en el mercado (Teensy 4.1, Adafruit). El microcontrolador de sincronización de tiempo transmitió la señal de sincronización de tiempo a la computadora de rastreo de los imanes a través de una placa adaptadora personalizada.

*Rastreo muscular móvil en diversas actividades*

Construimos un pasillo para guiar a los pavos a través de las actividades de terreno variable. Apilamos cajas pliométricas (Yes4All) en el pasillo a alturas de 20 cm, 41 cm y 61 cm para el ascenso y descenso vertical. Al finalizar las pruebas de ascenso y descenso vertical, colocamos rampas (Happy Ride Folding Dog Ramp, PetSafe) hacia arriba y hacia abajo de las cajas pliométricas con inclinaciones de 10° y 18° para que los pavos ascendieran y descendieran. Aparte de las actividades en terreno variable, permitimos a un pavo (Ave A) vagar libremente dentro de su recinto mientras seguíamos registrando la MM.

*Validación de la magnetomicrometría con pruebas de laboratorio para la actividad en terreno variable*

Para las pruebas de laboratorio, utilizamos superpegamento (Krazy Glue) para fijar cada una de dos imanes N48SH en dos placas redondas de LEGO de 1x1. Fijamos estas placas redondas de LEGO a un bloque técnico de LEGO de 1x6, uno en cada extremo, para separar el par de imanes a una distancia fija de 40 mm (véase la Figura Suplementaria 7A). Tomamos imágenes de este par de imanes utilizando la FM para validar la distancia fija de 40 mm entre ellas. Específicamente, después de la validación de la precisión de la MM de dos de las aves, recogimos datos de la FM con el par de imanes estáticamente en el volumen y utilizamos el promedio de los últimos tres segundos de cada una de estas dos colecciones de la FM para confirmar la distancia entre los dos imanes.

Para determinar el volumen de barrido de este par de imanes validado por la FM con una distancia de 40 mm durante las pruebas de laboratorio, analizamos los datos de rastreo de los imanes de todas las rampas y cambios de elevación verticales para determinar el rango completo de las posiciones tridimensionales de los imanes en relación con la serie de sensores de la MM en los tres pavos (véase la Figura Suplementaria 6).

Primero alineamos y centramos el par de imanes bajo la serie, y utilizamos las posiciones z de los imanes rastreados para colocar los imanes a la profundidad más cercana posible sin salirse del rango de detección a escala completa de los sensores (~1 cm). A continuación, utilizamos placas de LEGO de 3.2 mm de grosor para reforzar las profundidades restantes (véase la Figura Suplementaria 7A). En cada profundidad, barrimos manualmente desde el centro hacia fuera y hacia atrás a lo largo de los ejes x e y hasta el punto en el que el imán más lejano llegaba justo más allá de los requisitos del volumen de prueba derivados de la actividad del terreno variable (véase la Figura Suplementaria 7B).

*Análisis de datos*

Procesamos a posteriori los datos de la FM utilizando XMA Lab (Knörlein et al. 2016). Todos los datos de la FM y la MM se dejaron sin filtrar.

Alineamos los datos de la MM y la FM utilizando la señal de sincronización temporal e interpolamos linealmente los datos de la FM en los puntos temporales de medición de la MM. A continuación, debido a la imprecisión de la señal de sincronización temporal del sistema de rayos X, utilizamos la optimización local para alinear aún más las señales de la MM y la FM mientras interpolábamos iterativamente la FM. Durante un ensayo (una recolección de datos para un pavo a una velocidad), en el que la computadora de rastreo no recibió una sincronización temporal, utilizamos la optimización global para alinear las señales de la MM y la FM. Para validar el uso de la optimización global para la sincronización, probamos esta misma optimización global en todos los demás ensayos y encontramos que la optimización global localizó con éxito todas las señales de sincronización temporal.

Estimamos el ruido del procesamiento manual de la FM procesando independientemente un conjunto de diez ciclos de marcha tres veces (reprocesando manualmente los datos de vídeo dos veces sin referencia a los datos procesados anteriormente). A continuación, calculamos la varianza en cada punto de tiempo y utilizamos la raíz cuadrada de la varianza media como nuestra estimación del ruido del procesamiento manual de la FM (véase la Figura Suplementaria 4). Calculamos el ruido de la MM ajustado restando la varianza media del ruido de procesamiento manual de la FM de la varianza de la diferencia entre las señales de la MM y la FM para cada ave, y luego sacando la raíz cuadrada.

En la Figura 3, para los ciclos de la marcha en los que sólo era visible un golpe de la punta del pie en el vídeo, normalizamos el ciclo de la marcha utilizando el tiempo de las señales máximas de la MM en los ciclos de la marcha anterior y actual.

En la Figura Suplementaria 7, todos los datos se muestran en forma de gráfico de dispersión con una aproximación de spline cúbico de suavizado, utilizando un parámetro de suavizado de 0.8 (de Boor 1978). La desviación típica trazada se calculó como la raíz de la media cuadrática de los valores ajustados por el spline, también suavizados con un parámetro de suavización de 0.8.

## Conflicto de intereses

CT, SY y HH han presentado patentes sobre el concepto de magnetomicrometría titulado " Method for neuromechanical and neuroelectromagnetic mitigation of limb pathology " (patente WO2019074950A1) y sobre las estrategias de aplicación de la magnetomicrometría titulada " Magnetomicrometric advances in robotic control " (patente pendiente en EE.UU. 63/104942). El resto de los autores declaran que la investigación se llevó a cabo en ausencia de cualquier relación comercial o financiera que pudiera interpretarse como un conflicto de intereses potencial.

## Contribuciones de los autores

CT desarrolló la estrategia de magnetomicrometría, dirigió la concepción y el diseño experimental, asistió en las cirugías, realizó la recolección de datos, la documentación y el análisis, y dirigió la preparación del manuscrito. SY diseñó, validó y supervisó la fabricación del sistema integrado de detección del campo magnético, estableció el marco de medición en tiempo real habilitado por WiFi, contribuyó a la concepción y el diseño experimental, asistió en las cirugías, realizó la recolección de datos, la documentación y el análisis, y contribuyó a la preparación del manuscrito. WC, EC contribuyeron a la concepción y el diseño experimental, ayudaron en las cirugías, realizaron la recolección de datos, la documentación y el análisis, y contribuyeron a la preparación del manuscrito. MO contribuyó a la concepción y el diseño experimental, ayudó en las cirugías y contribuyó a la preparación del manuscrito. TR contribuyó a la concepción y el diseño experimental, dirigió las cirugías, asistió a la recolección y el análisis de datos, y contribuyó a la preparación del manuscrito. HH concibió la estrategia de magnetomicrometría, supervisó la financiación del proyecto, ayudó a la gestión general del estudio, contribuyó a la concepción y el diseño experimental, asistió al análisis de los datos y ayudó a la preparación del manuscrito.

## Financiación

## Agradecimientos

## Disponibilidad de datos

Todos los datos necesarios para respaldar las conclusiones del artículo se incluyen en el texto principal y en los materiales suplementarios. Además, los datos brutos de la magnetomicrometría y la fluoromicrometría procesados están disponibles a través de Zenodo en https://doi.org/10.5281/zenodo.6952783 (Taylor, Yeon, et al. 2022a).

## Referencias

Taylor et al.                                                                                                                  Rastreo Muscular MóvilGriffiths, RI. 1987. "Ultrasound Transit Time Gives Direct Measurement of Muscle Fibre Length in Vivo." *Journal of Neuroscience Methods* 21 (2–4): 159–65.

Hammond, P. 1960. "Electric and Magnetic Images." *The Institution of Electrical Engineers* 379: 306–13. https://doi.org/10.1.1.165.5581.

Knörlein, Benjamin J, David B Baier, Stephen M Gatesy, JD Laurence-Chasen, and Elizabeth L Brainerd. 2016. "Validation of XMALab Software for Marker-Based XROMM." *Journal of Experimental Biology* 219 (23): 3701–11.

Marsh, Richard L. 2016. "Speed of Sound in Muscle for Use in Sonomicrometry." *Journal of Biomechanics* 49 (16): 4138–41. https://doi.org/10.1016/j.jbiomech.2016.10.024.

Prilutsky, B I, W Herzog, and T L Allinger. 1996. "Mechanical Power and Work of Cat Soleus, Gastrocnemius and Plantaris Muscles during Locomotion: Possible Functional Significance of Muscle Design and Force Patterns." *Journal of Experimental Biology* 199 (4): 801–14. https://doi.org/10.1242/jeb.199.4.801.

Roberts, Thomas J, Richard L Marsh, Peter G Weyand, and C Richard Taylor. 1997. "Muscular Force in Running Turkeys: The Economy of Minimizing Work." *Science* 275 (5303): 1113–15.

Rosa, Luis G, Jonathan S Zia, Omer T Inan, and Gregory S Sawicki. 2021. "Machine Learning to Extract Muscle Fascicle Length Changes from Dynamic Ultrasound Images in Real-Time." *PloS One* 16 (5): e0246611.

Sikdar, Siddhartha, Qi Wei, and Nelson Cortes. 2014. "Dynamic Ultrasound Imaging Applications to Quantify Musculoskeletal Function." *Exercise and Sport Sciences Reviews* 42 (3): 126–35. https://doi.org/10.1249/JES.0000000000000015.

Tarantino, Sergio, Francesco Clemente, D Barone, Marco Controzzi, and CJSR Cipriani. 2017. "The Myokinetic Control Interface: Tracking Implanted Magnets as a Means for Prosthetic Control." *Scientific Reports* 7 (1): 1–11.

Taylor, Cameron R, Haley G Abramson, and Hugh M Herr. 2019. "Low-Latency Tracking of Multiple Permanent Magnets." *IEEE Sensors Journal* 19 (23): 11458–68.

Taylor, Cameron R, William H Clark, Ellen G Clarrissimeaux, Seong Ho Yeon, Matthey J Carty, Stuart R Lipsitz, Roderick T Bronson, Thomas J Roberts, and Hugh M Herr. 2022. "Clinical Viability of Magnetic Bead Implants in Muscle," *Frontiers in Bioengineering and Biotechnology*.

Taylor, Cameron R, Shriya S Srinivasan, Seong Ho Yeon, Mary Kate O'Donnell, Thomas J Roberts, and Hugh M Herr. 2021. "Magnetomicrometry." *Science Robotics* 6 (57): eabg0656.

Taylor, Cameron R., Seong Ho Yeon, William H. Clark, Ellen G. Clarrissimeaux, Mary Kate O'Donnell, Thomas J. Roberts, and Hugh M. Herr. 2022a. "Untethered Muscle Tracking Supplementary Materials." Zenodo. https://doi.org/10.5281/zenodo.6952783.

Taylor, Cameron R., Seong Ho Yeon, William H. Clark, Ellen G. Clarrissimeaux, Mary Kate O'Donnell, Thomas J. Roberts, and Hugh M. Herr. 2022b. "Untethered Muscle Tracking Using Magnetomicrometry." bioRxiv. https://doi.org/10.1101/2022.08.02.502527.

Taylor, Cameron R., Seong Ho Yeon, William H. Clark, Ellen G. Clarrissimeaux, Mary Kate O'Donnell, Thomas J. Roberts, and Hugh M. Herr. 2022c. "Untethered Muscle Tracking Using Magnetomicrometry." *Frontiers in Bioengineering and Biotechnology*.

Van Hooren, Bas, Panayiotis Teratsias, and Emma F. Hodson-Tole. 2020. "Ultrasound Imaging to Assess Skeletal Muscle Architecture during Movements: A Systematic Review of Methods, Reliability, and Challenges." *Journal of Applied Physiology* 128 (4): 978–99. https://doi.org/10.1152/japplphysiol.00835.2019.
16

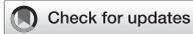

# Untethered muscle tracking using magnetomicrometry

Cameron R. Taylor[1†], Seong Ho Yeon[1†], William H. Clark[2], Ellen G. Clarrissimeaux[1], Mary Kate O'Donnell[2,3], Thomas J. Roberts[2*‡] and Hugh M. Herr[1*‡]

[1]K. Lisa Yang Center for Bionics, Massachusetts Institute of Technology, Cambridge, MA, United States, [2]Department of Ecology, Evolution, and Organismal Biology, Brown University, Providence, RI, United States, [3]Department of Biology, Lycoming College, Williamsport, PA, United States

Muscle tissue drives nearly all movement in the animal kingdom, providing power, mobility, and dexterity. Technologies for measuring muscle tissue motion, such as sonomicrometry, fluoromicrometry, and ultrasound, have significantly advanced our understanding of biomechanics. Yet, the field lacks the ability to monitor muscle tissue motion for animal behavior outside the lab. Towards addressing this issue, we previously introduced magnetomicrometry, a method that uses magnetic beads to wirelessly monitor muscle tissue length changes, and we validated magnetomicrometry *via* tightly-controlled *in situ* testing. In this study we validate the accuracy of magnetomicrometry against fluoromicrometry during untethered running in an *in vivo* turkey model. We demonstrate real-time muscle tissue length tracking of the freely-moving turkeys executing various motor activities, including ramp ascent and descent, vertical ascent and descent, and free roaming movement. Given the demonstrated capacity of magnetomicrometry to track muscle movement in untethered animals, we feel that this technique will enable new scientific explorations and an improved understanding of muscle function.

KEYWORDS

biomechanics, muscle tracking, implantable technology, wearable technology, motion tracking, magnetomicrometry, magnetic beads, magnet tracking

# Introduction

Muscle length measurements have driven important discoveries in movement biomechanics (Fowler et al., 1993), informed models of motor control (Prilutsky et al., 1996; Roberts et al., 1997), and provided strategies for prosthetic and robotic design (Eilenberg et al., 2010). For decades, sonomicrometry (SM) has informed how muscles move, providing high accuracy (70 μm resolution) and high bandwidth (>250 Hz) (Griffiths 1987). Fluoromicrometry (FM) expanded the muscle tracking toolkit, enabling high accuracy (90 μm precision) and high bandwidth (>250 Hz) for high-marker-count tracking (Brainerd et al., 2010; Camp et al., 2016). Further, image-based ultrasound (U/S) added the capability to non-invasively track muscle geometries (Fukunaga et al., 2001; Sikdar et al., 2014; Clark and Franz, 2021).

Yet, collecting direct muscle length measurements in natural environments remains infeasible, and thus indirect muscle length estimation is still used for observing natural movements. For instance, muscle lengths are estimated using joint angles *via* biophysical models (Delp et al., 2007). These approximations are used due to the limitations of current muscle motion sensing techniques, all of which are tethered or bulky. SM and U/S both require tethered connections to bulky hardware for sensing (Biewener et al., 1998; Clark and Franz, 2021), with SM requiring advanced surgery and percutaneous wires. And while FM does not require a tethered connection, it is limited to a volume approximately the size of a soccer ball, requires equipment the size of a small room, and is time-constrained due to thermal limitations and subject radiation exposure (Brainerd et al., 2010).

Present muscle length tracking technologies also require substantial post-processing time, hindering their use in longitudinal studies. SM requires accounting for and filtering out artifacts such as triggering errors (Marsh 2016), FM requires point labeling in stereo images (Brainerd et al., 2010), and U/S requires fascicle labeling (Van Hooren et al., 2020), all of which require at least some manual processing. While machine learning techniques have shown potential for automatic fascicle length tracking from ultrasound images, the current lack of reliability in tracking cross-activity measurements ($R^2$ = 0.05 for single-subject cross-activity training of a support vector machine) prevents such a strategy from being applicable toward sensing fascicle lengths during natural movement (Rosa et al., 2021).

Researchers need a sensing platform that can operate untethered in natural environments, sensing the full dynamic range of muscle movement in context. To address this need, we developed magnetomicrometry (MM), a minimally-invasive strategy for portable, real-time muscle tracking. MM uses an array of magnetic field sensors to locate and calculate the distance between two implanted magnetic beads with sub-millisecond time delay. This distance provides a measurement of the muscle tissue length between the implanted beads. MM allows continuous recording over an indefinite collection interval extending across hours, with the potential for continuous use across days, weeks, or years.

In prior work, we validated the MM concept against FM *via* tightly controlled *in situ* tests (Taylor et al., 2021). However, it has not previously been empirically demonstrated that the MM technique is robust for recording during untethered locomotion. In the present study we address this question. We investigate the robustness of MM during untethered activity that exhibits soft tissue artifacts (i.e., movement of the magnetic field sensors relative to the muscle) and changes in the relative orientation of the ambient magnetic field.

Herein we present MM as a robust, practical, and effective strategy for measuring muscle tissue length in an untethered freely-moving animal model. We first apply this technique to turkeys running on a treadmill and compare MM to FM to determine the method's accuracy. We then further investigate the use of MM to track muscle tissue length in freely-moving animals during ramp ascent and descent, vertical ascent and descent, and free roaming movement. We hypothesize that muscle tissue lengths during untethered motion can be tracked *via* MM with submillimeter accuracy and a strong correlation ($R^2$ > 0.5) to FM. Our validation of this tool in a mobile context enables tracking and investigation of muscle physiology in settings previously inaccessible to biomechanics researchers.

# Results

## Accuracy validation of magnetomicrometry against fluoromicrometry

To verify MM tracking accuracy during untethered activity, we tracked implanted magnetic bead pairs in turkey gastrocnemius muscles (right leg, three turkeys) using both MM and FM while the turkeys walked and ran at multiple speeds on a treadmill (see Figure 1 for the setup and tracking results, see Supplementary Figure S1 for a 3-D scan of the MM sensing array).

We compared the distances between the magnetic bead positions as measured by MM with their distances as measured by FM to evaluate accuracy during the treadmill activity (see Figure 2). The coefficients of determination ($R^2$ values) between MM and FM were 0.952, 0.860, and 0.967 for Birds A, B, and C, respectively (see also Supplementary Figure S2). The differences between MM and FM were −0.099 ± 0.186 mm, −0.526 ± 0.298 mm, and −0.546 ± 0.184 mm for Birds A, B, and C, respectively (see Supplementary Figure S3).

To determine the study-specific reliability of the manual FM processing (marker position labeling in the X-ray video data), ten gait cycles of raw FM data were independently manually relabeled three times for one bird at one speed. Across these three labelings for these ten gait cycles, manual FM processing was consistent to a standard deviation of 0.098 mm (see Supplementary Figure S4 for more details).

MM's 99-th percentile tracking time delays were 0.698 ms, 0.690 ms, and 0.664 ms for Birds A, B, and C, respectively (see also Supplementary Figure S5), and the MM data did not require any post-processing. In contrast, post-processing the FM data into marker-to-marker distances required approximately 84 manual processing hours spread across multiple months.

## Untethered muscle tracking across various activities

To investigate the feasibility of using MM during dynamic, natural motion, we constructed a series of obstacles for the turkeys to navigate. Specifically, we provided the turkeys with two ramp inclines (10° and 18°, see Figure 3) and three vertical elevation changes (20 cm, 41 cm, and 61 cm, see Figure 4). Because the purpose of these activities was to explore the range of dynamic motions that could be captured, we did not train the birds to navigate the ramps or vertical elevation changes repetitively, and thus variability is expected within the repeated tasks.

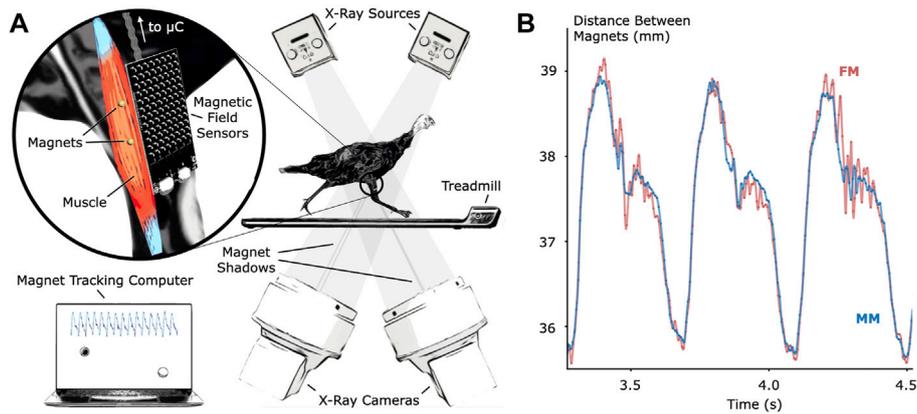

**FIGURE 1**
Validation of Untethered Muscle Tracking using Magnetomicrometry. **(A)** A magnetic field sensing array on the surface of the leg tracks the positions of two magnetic beads implanted into the muscle. A feather microcontroller (μC) in the turkey feathers wirelessly transmits the magnetic field data to a magnet tracking computer that calculates and displays the magnetomicrometry (MM) signal in real time. The turkeys walked and ran on a treadmill while x-ray video cameras recorded synchronized fluoromicrometry (FM) data for post-processing. **(B)** Comparison of MM (blue) with FM (red) to validate the MM accuracy. These representative results during running gait show the submillimeter accuracy of MM during untethered muscle length tracking.

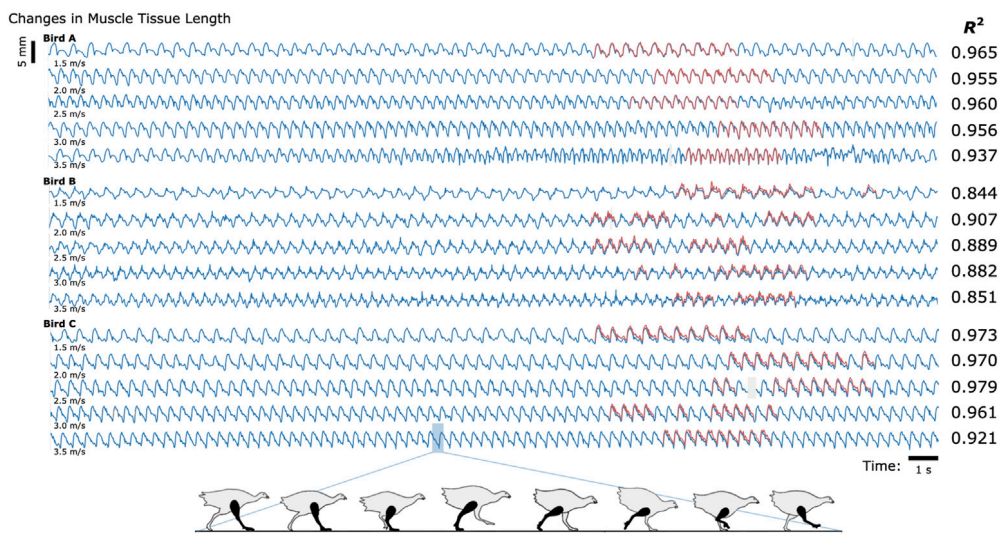

**FIGURE 2**
Untethered Muscle Tracking During Treadmill Running: Magnetomicrometry Versus Fluoromicrometry. Changes in muscle tissue length measured by MM (blue) and FM (red) for three turkeys at five speeds (30 s shown for each speed). The column to the right of the plots gives the coefficients of determination ($R^2$) between magnetomicrometry and fluoromicrometry corresponding to each turkey and speed. Gaps in the fluoromicrometry data are due to researcher selection of full gait cycles during which both magnetic beads were visible in both x-ray images. Gaps in the magnetomicrometry data (gray) are due to packet drops during wireless transmission of the magnetic field signals to the tracking computer (gaps below 50 ms interpolated in gray, gaps above 50 ms highlighted in gray). The turkey gait diagram below the plots shows the corresponding gait phases over one gait cycle.

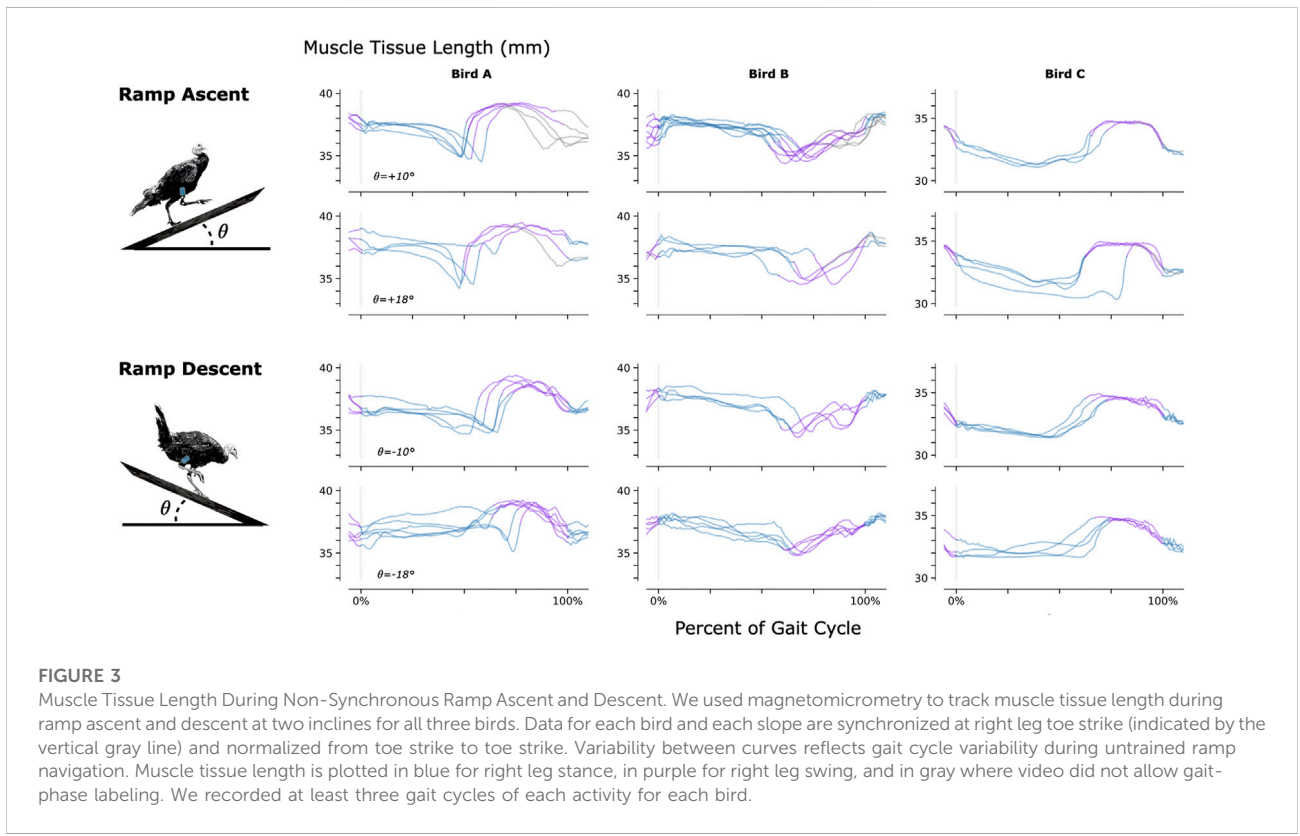

FIGURE 3
Muscle Tissue Length During Non-Synchronous Ramp Ascent and Descent. We used magnetomicrometry to track muscle tissue length during ramp ascent and descent at two inclines for all three birds. Data for each bird and each slope are synchronized at right leg toe strike (indicated by the vertical gray line) and normalized from toe strike to toe strike. Variability between curves reflects gait cycle variability during untrained ramp navigation. Muscle tissue length is plotted in blue for right leg stance, in purple for right leg swing, and in gray where video did not allow gait-phase labeling. We recorded at least three gait cycles of each activity for each bird.

For a video of one of the turkeys (bird A) navigating all of these obstacles with real-time MM data shown, see Supplementary Movie S1.

To further validate the accuracy of MM used during navigation of ramps and vertical elevation changes, we analyzed the magnetic bead tracking data from these activities to find the range of the tracked three-dimensional magnetic bead positions (see Supplementary Figure S6). We then affixed two magnetic beads 40 mm apart, validated the distance between them using FM (40.000 ± 0.017 mm), and swept this FM-validated magnetic bead pair under the MM sensing array through a volume exceeding these ranges (see Supplementary Figure S7). We monitored deviations from 40 mm in the MM signal during these benchtop tests and found a 99-th percentile error ($e_{99\%}$) of 1.000 mm (rounded up to the nearest micrometer).

Finally, to explore whether untethered muscle tracking *via* MM is viable in a fully free roaming context, we tracked muscle tissue length while one turkey (Bird A) roamed freely about its enclosure. The results of this data collection are shown in Figure 5.

## Discussion

We find that MM enables untethered muscle tissue length tracking with high correlation to FM ($R^2$ of 0.952, 0.860, and 0.967 for Birds A, B, and C, respectively) and submillimeter accuracy (AVG ± SD of -0.099 ± 0.186 mm, -0.526 ± 0.298 mm, and −0.546 ± 0.184 mm for Birds A, B, and C, respectively). These findings enable tracking and investigation of muscle contractile behavior in settings previously inaccessible to biomechanics researchers.

## Accuracy validation

The standard we used here to assess the accuracy of muscle length tracking using MM was FM. For magnets implanted superficially in muscles (at depths less than 2 cm), MM exhibits less noise than FM, but FM has the advantage of higher accuracy, especially at greater tissue depths (tracking depths in this study ranged from 11.2 mm to 26.6 mm). Indeed, our tests showed that for unobscured markers moving through the X-ray volume, FM was accurate to 0.030 mm. However, we note that marker tracking noise was a challenge for FM in this particular study due to the use of a large animal and the presence of hardware (the MM sensing array) that regularly obscured the markers during the tracking. These factors resulted in substantial manual labeling noise in the FM signal of 0.098 mm, instead of the 0.030 mm noise we found in our FM accuracy test, affecting the accuracy standard deviations reported above. Accounting for this manual labeling noise gives adjusted accuracy standard deviations of 0.158, 0.281, and 0.156 mm, for Birds A, B, and C, respectively (see Supplementary Figure S3).

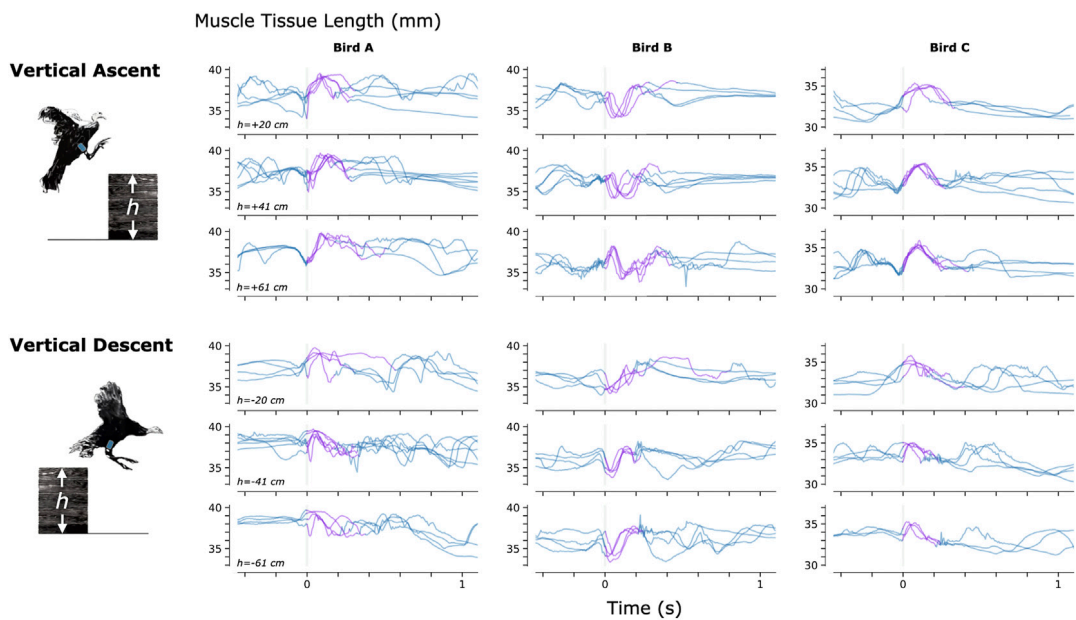

FIGURE 4
Muscle Tissue Length During Non-Synchronous Vertical Ascent and Descent. We used magnetomicrometry to track muscle tissue length during vertical ascent and descent at three heights for all three birds. Data for each bird and each height are synchronized at right leg toe-off (start of the aerial phase, indicated by the vertical gray line). Variability between curves reflects movement variability during untrained vertical ascent and descent. Muscle tissue length during contact with the ground is plotted in blue, and muscle tissue length during the aerial phase is plotted in purple. All data are shown, including scenarios in which significant wing-flapping occurred during jump up or down. We captured at least three recordings of each activity for each bird.

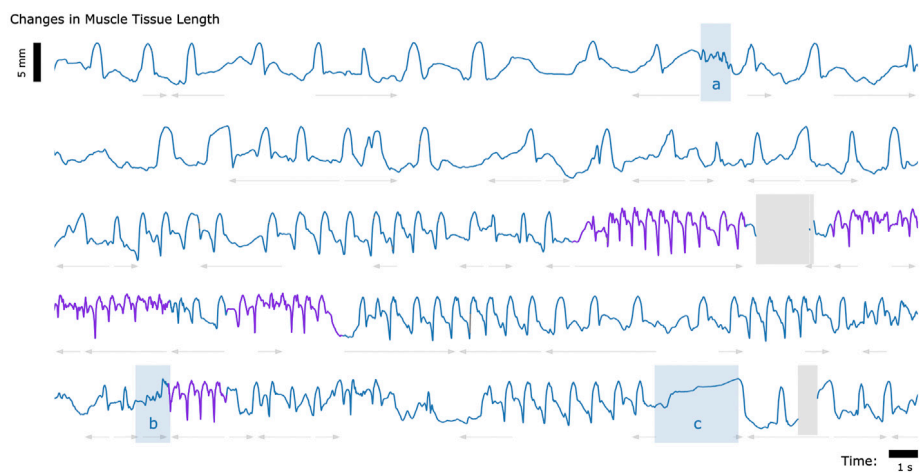

FIGURE 5
Muscle Tissue Length During Free Roaming Movement. Magnetomicrometry data was continuously collected for 150 s during free roaming activity. Muscle tissue length is plotted in blue during standing and walking and plotted in purple during running. Blue highlighted regions indicate muscle tissue length during (a) feather ruffling, (b) jumping, and (c) balancing on one leg. Gray arrows indicate when the turkey was turning left (left arrows) or turning right (right arrows). Gaps due to wireless transmission packet drops are shown in gray, as described in Figure 2.

Constraints to imaging volume make FM impractical during large-animal variable terrain activity, so we performed retrospective benchtop accuracy testing to further validate the MM data collected during navigation of the ramps and vertical elevation changes (see Supplementary Figure S7). Soft tissue artifacts during dynamic movements, such as tissue deformation or movement, result in depth and position changes of the magnetic beads relative to the MM sensing array. The benchtop tests investigated the accuracy of the MM measurements across the range of depths and positions of the magnetic beads that we observed during those activities (see Supplementary Figure S6). The error we observed in the benchtop tests ($e_{99\%}$ < 1 mm) was acceptable in comparison with the magnitude of the muscle contractions we observed during the variable terrain activity (average MM signal magnitude was 4.5 mm peak-to-peak). This suggests that MM robustly tracked the muscle tissue lengths during the variable terrain activities, despite any soft tissue artifacts that may have occurred during the dynamic movements required by those activities. These tests, however, highlight the importance of sensor placement. Higher accuracy is achieved when the MM sensing array is properly placed–centered over the implanted beads. MM with perfect magnetic field sensing would, in theory, be unaffected by movement of the board relative to the implanted beads, but the errors we observed suggest that the sensors are nonlinear. Magnet tracking nonlinearity compensation (e.g., *via* sensor calibration or three-dimensional sensor geometries) is thus an important area for future research. Meanwhile, in future work, larger sensing arrays with broader coverage would be advantageous to mitigate the need for careful placement of the array.

## Ambient magnetic fields

The software-based magnetic disturbance compensation we employed here (Taylor et al., 2019) was sufficient to compensate for ambient magnetic fields during untethered muscle tracking in the presence of large hydraulic ferromagnetic lift tables, a large, active treadmill motor, and a room full of active X-ray equipment. However, our uniform disturbance compensation strategy may be insufficient for the exceptional situation where a large ferromagnetic object is immediately adjacent to (within a few centimeters of) the tracked muscle. Thus, software-based compensation for spatially-non-uniform ambient magnetic fields may still be a valuable direction for future work to extend the robustness of MM to that potential scenario. Alternatively, ferromagnetic shielding could be used to physically perform disturbance compensation (Tarantino et al., 2017), but the shield would need to be sufficiently far away to prevent it from acting like a magnetic mirror, creating "image" magnets that would need to be tracked as well (Hammond 1960). Further, effective shielding would need to be thick enough to redirect most or all magnetic field disturbances, presenting a trade-off between the weight and the efficacy of the shielding.

## Range of behaviors

Figures 3–5 provide a sample of the range of behaviors that can be tracked using magnetomicrometry. Consistency in the curves was not strived for, expected, or desired. Rather, we intentionally preserved anomalous events in those data, such as single or multiple wing flaps during vertical ascent and descent and variable speed during ramp navigation, to explore the range of motor activities during which we could track the muscle activity.

## Applications

MM has the potential to work across scales (see Figure 6), from the ability to track both full-body and muscle movement of small organisms to the ability to track large magnetic beads implanted deep into large animal models. Mathematically, if the number of sensors is fixed and all system dimensions are scaled, the error as a percent of scaled magnetic bead excursion will remain unchanged (Taylor et al., 2019). However, larger sensing arrays can in principle be used when tracking very small or very large animals, resulting in an increase in tracking accuracy at those extremes. For instance, when tracking small animal muscle tissue, additional sensors can be embedded into the animal's environment, and when tracking large animal muscle tissue, the increased animal size accommodates the mounting of additional sensors to the animal. Thus, context-specific magnetic bead tracking systems can take advantage of the unique geometries afforded at each scale.

Not only can MM be used across size scales, but across time scales as well. Because there is no need for post-processing, MM data can be collected continuously, enabling the potential for longitudinal studies, including investigations into mechanisms such as neural degradation or plasticity over time.

MM's muscle length and velocity signals are different from, and complementary to, the signals from electromyography (EMG). While EMG provides a measure of muscle activation, which results more directly from neural commands, muscle length and velocity give information on the shape of the muscle, which in turn can refine our understanding of muscle physiology during a given motion task. Indeed, the combination of MM and EMG will allow for increased physiological understanding in new contexts where animals are in their natural environments.

In parallel work, we also demonstrate the viability of magnetic bead implants for human use, verifying comfort,

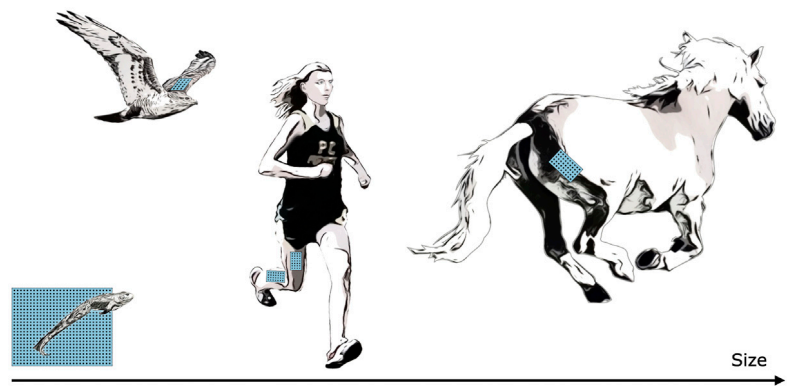

**FIGURE 6**
Muscle Tracking Across Scales. By changing the size of the magnetic field sensing array, we can track the distance between magnets at closer or farther distances, allowing us in principle to track muscle tissues at a range of scales, including frogs, hawks, persons, horses, or other animals. For small animals, such as the frog shown at bottom left, a fixed array below or beside the animal could track both the position of the animal and the muscle tissue length.

lack of implant migration, and biocompatibility (Taylor, Clark, et al., 2022). Due to the untethered nature of MM, this technique has applications in prosthetic and exoskeletal control. In its most straightforward implementation, a motor controller could directly control a robotic joint using the distance between two beads in each muscle of a flexor-extensor pair. However, the ability for MM to track additional muscles and to work in combination with EMG enables a range of new strategies for human-machine interfacing.

## Limitations

In this study, we implanted the magnetic beads approximately 3.5 cm away from one another, based on previous work (Taylor et al., 2021), to ensure that the magnetic beads would not migrate toward one another. If smaller or larger (or differently shaped) magnetic bead implants are used (for instance, in a smaller or larger animal model), the effect of separation distance on stability against migration would need to be re-investigated for the different sizes (and different magnetization strengths) of the implants.

For the benchtop accuracy validation tests, we assumed that the tracked bead positions were a good approximation for the true bead positions. We used the magnetic bead position tracking information from the variable terrain MM data to determine the boundaries of the volume to test. Then, at the start of the tests, we used tracked bead positions to locate the centered, minimum-depth position (the closest position within the full scale range of the sensors), then used blocks of known dimensions to sweep through the benchtop-emulated tissue depth and enforce the volume boundaries. Noting that MM was accurate to within a millimeter throughout the volume, we found these assumptions reasonable for these tests.

As for any muscle tissue tracking, the location of the implanted tracking devices will determine the length measured. For studies where the aim is to relate measured length changes to muscle contractile properties (e.g., length-tension or force-velocity relationships), it is essential that the markers are aligned along the fascicle axis. In the present study we embedded magnets at locations in the turkey muscles that would ensure the magnets stayed in place over a period of months, and at depths that were favorable for sensor function. Thus, patterns of length change do not directly represent patterns of muscle fascicle length change and can be influenced significantly by dynamic changes in muscle architecture during contraction. This is reflected in the opposite muscle tissue length changes seen during the swing phase of Bird B relative to Birds A and C during ramp navigation (see Figure 3). MM, FM, and SM all suffer from this same issue, and thus for any of these techniques, careful surgical placement is warranted.

## Sensing improvements

The suite of electronics for MM is immediately upgradeable as new industry standards develop. The tracking system benefits from global developments in low-cost magnetic field sensors due to the widespread manufacturing of inertial measurement units for devices such as cell phones, video game controllers, and autonomous vehicles. Continuing improvements in magnetic field sensors, capacitors, and microcontrollers will cause direct improvements to the accuracy, efficiency and speed of the tracking system and will allow the tracking of even smaller implants at greater depths.

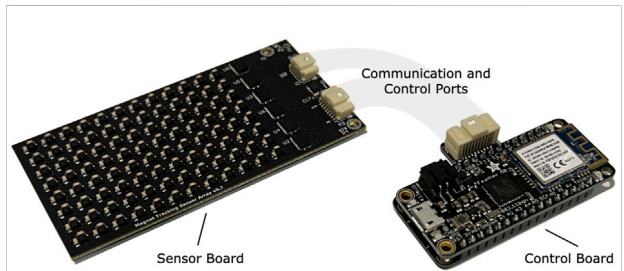

**FIGURE 7**
Magnetomicrometry Embedded System. We fabricated a custom sensor board (left) and a custom control board (right) for this study. The sensor board holds the magnetomicrometry sensing array, consisting of 96 magnetic field sensors arranged with a spacing of 5.08 mm. Digital multiplexers on the sensor board allow time-domain multiplexing, enabling a single microcontroller on the control board to communicate with and control all magnetic field sensors on the sensor board. The control board merges the data from the sensor board and streams the data wirelessly to the magnet tracking computer. The sensor board and control board weigh 24 g and 12 g, respectively.

## Summary

Here, we demonstrate the use of MM for untethered muscle tracking. We validate, against FM, the submillimeter accuracy of MM in an awake, active turkey model ($R^2 \geq 0.860$, $\mu \leq 0.546$ mm, $\sigma \leq 0.298$ mm) with a real-time computing time delay of less than a millisecond ($\eta_{0.99} \leq 0.698$ ms). We further demonstrate the use of MM in untethered muscle tracking during ramp ascent and descent, vertical ascent and descent, and free roaming movement. These results encourage the use of MM in future biomechanics investigations, as well as in prosthetic and exoskeletal control. We hope that MM will enable a variety of new experiments and technologies, and we look forward to the further development and application of this technology.

## Methods

All animal experiments were approved by the Institutional Animal Care and Use Committees at Brown University and the Massachusetts Institute of Technology. Wild turkeys (*Meleagris gallopavo*, adult female) were obtained from local breeders and maintained in the Animal Care Facility at Brown University on an *ad libitum* water and poultry feed diet. We used three animals in this study.

### Surgical procedure

One pair of 3-mm-diameter Parylene-coated magnetic beads (N48SH) were implanted into the right lateral gastrocnemius muscle of each turkey, with a target magnetic bead separation distance of 3.5 cm. For details on the surgical procedure and implants, see Taylor et al., 2022. A 1-month recovery period was given before the start of the data collection.

### Magnetomicrometry

For this study, we designed a custom magnetic field sensing array (see Figure 7). The sensing array was equipped with 96 magnetic field sensors (LIS3MDL, STMicroelectronics) spaced 5.08 mm apart in an 8-by-12 grid. Each sensor was supplied with nonmagnetic capacitors (VJ1206Y105KCXAT and VJ0603Y104KCXAT, Vishay). Seven digital multiplexers on the sensing array allowed time-domain multiplexing (one 74HC138BQ,115 multiplexing into six 74HC154BQ,118, Nexperia) *via* a wired connection. The sensing array was connected through a custom adapter board to an off-the-shelf wireless microcontroller embedded system (Feather M0 WiFi microcontroller, Adafruit), which was powered by a lithium-ion polymer battery (3.7 V, 1800 mA·h, 29 g). The microcontroller sampled the magnetic field signals at 155 Hz and wirelessly transmitted them to the magnet tracking computer *via* a WiFi router (Nighthawk R6900P, Netgear). The tracking algorithm ran in real-time on the magnet tracking computer, a Dell Precision 5550 laptop (Ubuntu 20.04 operating system) with 64 GB of random-access memory and an Intel i7 8-Core Processor, running at 2.30 GHz. The tracking algorithm used, including the strategy for disturbance compensation, is fully-detailed in previous work (Taylor et al., 2019).

We affixed the sensing array to the limb using Opsite Flexifix adhesive film (Smith & Nephew). To secure the array, we first applied a base layer of the adhesive film to the skin. We then positioned the sensing array over the leg and wrapped the adhesive film around the sensing array and the leg. To maintain a sufficient minimum distance between the magnetic beads and the sensing array, we positioned layers of foam between the base adhesive and the array. We then secured the control board and battery within the back feathers of the turkey.

### Accuracy validation of magnetomicrometry against fluoromicrometry

We used the W. M. Keck Foundation XROMM Facility at Brown University (Brainerd et al., 2010) to perform FM. We collected X-ray video from two intersecting X-ray beams oriented at 51 degrees relative to one another. We mounted a treadmill with a wooden base (TM145, Horizon Fitness) between the X-ray cameras and the X-ray sources and built a housing over the treadmill with a movable wall to position the birds within the capture window.

The turkeys walked and ran at five speeds (1.5, 2.0, 2.5, 3.0, and 3.5 m/s) in a randomized order until at least ten gait cycles were visible within the FM capture volume for each speed. We collected FM data at 155 Hz.

Time syncing was performed *via* a coaxial cable connection from FM to an off-the-shelf microcontroller development board (Teensy 4.1, Adafruit). The time-syncing microcontroller relayed the time sync signal to the magnet tracking computer *via* a custom adapter board.

## Untethered muscle tracking across various activities

We constructed a hallway to guide the turkeys through the variable terrain activities. We stacked plyometric boxes (Yes4All) in the hallway to heights of 20 cm, 41 cm, and 61 cm for vertical ascent and descent. Upon completion of the vertical ascent and descent tests, we placed ramps (Happy Ride Folding Dog Ramp, PetSafe) up to and down from the plyometric boxes at inclines of 10° and 18° for the turkeys to ascend and descend. Separate from the variable terrain activities, we then allowed one turkey (Bird A) to roam freely within its enclosure while we continued to record MM.

## Benchtop magnetomicrometry validation for the variable terrain activity

For benchtop testing, we used super glue (Krazy Glue) to affix each of two N48SH magnetic beads into two 1 × 1 round LEGO plates. We attached these round LEGO plates to a 1 × 6 LEGO technic block, one at each end, to separate the pair of magnetic beads by a fixed distance of 40 mm (see Supplementary Figure S7A). We imaged this pair of beads using FM to validate the 40 mm fixed distance between them. Specifically, after the MM accuracy validation of two of the birds, we collected FM data with the magnet pair statically in the volume and used the average of the last 3 seconds from each of these two FM collections to confirm the distance between the two beads.

To determine the volume to sweep this FM-validated 40-mm-distanced bead pair during benchtop testing, we analyzed the magnetic bead tracking data from all ramps and vertical elevation changes to determine the full range of the three-dimensional magnetic bead positions relative to the MM sensing array across all three turkeys (see Supplementary Figure S6).

We first aligned and centered the magnet pair under the array, and we used the tracked magnet z-positions to place the magnets at the closest depth that was possible while still within the full-scale sensing range of the sensors (~1 cm). We then used 3.2-mm-thick 1 × 6 LEGO plates to enforce the remaining depths (see Supplementary Figure S7A). At each depth, we manually swept from center out and back along the *x* and *y* axes to the point where the farthest magnet reached just beyond the test volume requirements derived from the variable terrain activity (see Supplementary Figure S7B).

## Data analysis

We post-processed the FM data using XMA Lab (Knörlein et al., 2016). All FM and MM data were left unfiltered.

We aligned the MM and FM data using the time sync signal and linearly interpolated the FM data at the MM measurement time points. Then, due to imprecision of the time sync signal from the X-ray system, we used local optimization to further align the MM and FM signals while iteratively interpolating FM. During one trial (one data collection for one turkey at one speed), where the tracking computer did not receive a time sync, we used global optimization to align the MM and FM signals. To validate the use of global optimization for synchronization, we tested this same global optimization on all other trials and found that the global optimization successfully located all time sync signals.

We estimated the noise from manual FM processing by independently processing one set of ten gait cycles three times (manually re-processing the video data twice without reference to the previously processed data). We then calculated the variance at each time point and used the square root of the average variance as our estimate of the FM manual processing noise (see Supplementary Figure S4). We calculated the adjusted MM noise by subtracting the average variance of the FM manual processing noise from the variance of the difference between the MM and FM signals for each bird, then taking the square root.

In Figure 3, for gait cycles where only one toe strike was visible in the video, we normalized the gait cycle using the timing of the peak MM signals in the previous and current gait cycles.

In Supplementary Figure S7, all data are shown plotted as a scatterplot with a smoothing cubic spline approximation, using a smoothing parameter of 0.8 (Boor, 1978). The plotted standard deviation was calculated as the root-mean-square of the spline-adjusted values, also smoothed with a smoothing parameter of 0.8.

## Data availability statement

All data needed to support the conclusions in the paper are included in the main text and Supplementary Materials. In addition, the raw magnetomicrometry and processed fluoromicrometry data is available *via* Zenodo at https://doi.org/10.5281/zenodo.6952783 (Taylor et al., 2022a).

## Ethics statement

The animal study was reviewed and approved by the Institutional Animal Care and Use Committees at Brown University and the Massachusetts Institute of Technology.

## Author contributions

CT developed the magnetomicrometry strategy, led the experimental conception and design, assisted in surgeries, performed data collection, documentation, and analysis, and led the manuscript preparation. SY designed, validated, and oversaw the fabrication of the magnetic field sensing embedded system, set up the WiFi-enabled real-time measurement framework, contributed to experimental conception and design, assisted in surgeries, performed data collection, documentation, and analysis, and contributed to manuscript preparation. WC, EC contributed to experimental conception and design, assisted in surgeries, performed data collection, documentation, and analysis, and contributed to manuscript preparation. MO contributed to experimental conception and design, assisted in surgeries, and contributed to manuscript preparation. TR contributed to experimental conception and design, led the surgeries, assisted in data collection and analysis, and contributed to manuscript preparation. HH conceived the magnetomicrometry strategy, oversaw project funding, assisted with general study management, contributed to experimental conception and design, assisted in data analysis, and aided in manuscript preparation.

## Funding

This work was funded by the Salah Foundation, the K. Lisa Yang Center for Bionics at MIT, the MIT Media Lab Consortia, NIH grant AR055295, and NSF grant 1832795.

## Acknowledgments

The authors thank, non-exhaustively, Elizabeth Brainerd, Andreas Burger, Ziel Camara, John Capano, Matthew Carty, Tyler Clites, Charlene Condon, Kathy Cormier, Jimmy Day, Bruce Deffenbaugh, Jennie Ehlert, Wendy Ehlert, Rachel Fleming, Lisa Freed, Robert Gnos, Deborah Grayeski, Alan Grodzinsky, Samantha Gutierrez-Arango, Ayse Guvenilir, Kale Hansen, Mallory Hansen, Guillermo Herrera-Arcos, Crystal Jones, Kylie Kelley, Duncan Lee, Aimee Liu, Richard Marsh, Richard Molin, Chad Munro, Paris Myers, Jarrod Petersen, Mitchel Resnick, Lindsey Reynolds, Andy Robinson, Jacob Rose, Amy Rutter, Shriya Srinivasan, Erika Tavares, Sara Taylor and Beni Winet for their helpful advice, suggestions, feedback, and support. Inclusion in this list of acknowledgments does not indicate endorsement of this work. A preprint of this manuscript was deposited on bioRxiv (Taylor, Yeon, et al., 2022b).

## Conflict of interest

CT, SY, and HH have filed patents on the magnetomicrometry concept entitled "Method for neuromechanical and neuroelectromagnetic mitigation of limb pathology" (patent WO2019074950A1) and on implementation strategies for magnetomicrometry entitled "Magnetomicrometric advances in robotic control" (US pending patent 63/104942).

The remaining authors declare that the research was conducted in the absence of any commercial or financial relationships that could be construed as a potential conflict of interest.

## Publisher's note

All claims expressed in this article are solely those of the authors and do not necessarily represent those of their affiliated organizations, or those of the publisher, the editors and the reviewers. Any product that may be evaluated in this article, or claim that may be made by its manufacturer, is not guaranteed or endorsed by the publisher.

## Supplementary material

The Supplementary Material for this article can be found online at: https://www.frontiersin.org/articles/10.3389/fbioe.2022.1010275/full#supplementary-material